\documentclass[preprint]{aastex61}
\usepackage{amsmath}
\usepackage{mathrsfs}

\newcommand\aastex{AAS\TeX}

\received{August 29, 2018}
\revised{January 7, 2019}
\accepted{\today}
\submitjournal{ApJS}

\shorttitle{\aastex\ sample article}
\shortauthors{Li and Chen}

\begin{document}

\title{Simulating Non-hydrostatic atmospheres on Planets (SNAP): formulation, validation and application to the Jovian atmosphere}

\correspondingauthor{Cheng Li}
\email{cli@gps.caltech.edu}

\author[0000-0002-8280-3119]{Cheng Li}
\affil{California Institute of Technology, 1200 E California Blvd, Pasadena, CA, 91106}
\author[0000-0002-9952-9048]{Xi Chen}
\affil{Geophysical Fluid Dynamics Laboratory, NOAA}
\affil{Princeton University, Princeton, NJ, 08544}



\begin{abstract}

A new non-hydrostatic and cloud-resolving atmospheric model is developed for studying moist convection and cloud formation in planetary atmospheres. It is built on top of the Athena++ framework, utilizing its static/adaptive mesh-refinement, parallelization, curvilinear geometry, and dynamic task scheduling. We extend the original hydrodynamic solver to vapors, clouds, and precipitation. Microphysics is formulated generically so that it can be applied to both Earth and Jovian planets. We implemented the Low Mach number Approximate Riemann Solver (LMARS) for simulating low speed atmospheric flows in addition to the usual Roe and HLLC Riemann solvers. Coupled with a fifth-order Weighted Essentially Nonoscillatory (WENO) subgrid-reconstruction method, the sharpness of critical fields such as clouds is well-preserved, and no extra hyperviscosity or spatial filter is needed to stabilize the model. Unlike many atmospheric models, total energy is used as the prognostic variable of the thermodynamic equation. One significant advantage of using total energy as a prognostic variable is that the entropy production due to irreversible mixing process can be properly captured. The model is designed to provide a unified framework for exploring planetary atmospheres across various conditions, both terrestrial and Jovian. First, a series of standard numerical tests for Earth's atmosphere is carried out to demonstrate the performance and robustness of the new model. Second, simulation of an idealized Jovian atmosphere in radiative-convective equilibrium shows that 1) the temperature gradient is superadiabatic near the water condensation level because of the changing of the mean molecular weight, and 2) the mean profile of ammonia gas shows a depletion in the subcloud layer down to nearly 10 bars. Relevance to the recent Juno observations is discussed.
%
%
%
%
%
%
%
%

\end{abstract}


\section{Introduction} \label{sec:intro}
Atmospheric science for Earth and planets has come to a point where a unified model is needed to study diverse climates on numerous exotic worlds. Earth's atmosphere has been observed and studied extensively. Yet, from a broader perspective, the possible range of Earth's climate only occupies a niche of the parameter space of all planetary atmospheres that have been observed in the universe. Mars, Pluto, and Triton have atmospheres made up mostly of condensible gases. Condensation of CO$_2$ or N$_2$ on to the planetary surface removes a significant portion of the total mass of the atmosphere, resulting in a global circulation driven by sublimation and condensation \citep{Pierrehumbert16,Ding18}. The atmosphere of giant planets harbors multiple condensible species whose molecular weights are much larger than the ambient hydrogen atmosphere. The loading of heavy molecules stratifies the atmosphere and inhibits moist convection \citep{Sugiyama14,Li15,Leconte17,Friedson17}. Beyond the solar system, nearly 3000 exoplanets have been discovered at the time of writing\footnote{Statistics are gathered from www.exoplanets.org}. Many of them have no counterparts in our solar system (hot Jupiters, super Earth, lava planets, etc.). It is, therefore, challenging to picture their atmospheric circulation pattern using the knowledge from studying the solar system.

Global circulation models (GCM) are powerful tools for simulating atmospheric flows on planets. Limited by computational resources, the horizontal resolution of a GCM is usually around 100 kilometers for Earth (e.g. AM4 model: \citealt{Zhao18,Lin04}) and 500 kilometers for Jovian planets (e.g. Jupiter GCM: \citealt{Schneider09}). Models with such a low resolution cannot resolve small-scale processes like moist convection or cloud formation. Therefore, they must employ parameterization schemes that approximate the effect of small-scale processes. The parameterization schemes for Earth are usually motivated by observations or by regional numerical models that resolve these processes (see \citealt{Arakawa11} for a review). For planets other than Earth, the absence of granular observations precludes the possibility of using observations to constrain the parameterization scheme. Direct modeling of convection embedded in a large-scale environment becomes necessary to understand how the parameterization scheme should be for a planet GCM. Right now, many planet GCMs either borrow parameterization schemes designed for Earth (e.g. \citealt{Schneider09}), or neglect any parameterization of convection (e.g. \citealt{Lian10}). Such approaches may be valid for investigating problems like jet formation or global temperature distribution, but it is doubtful when one tries to use a low-resolution GCM to understand how tracers or clouds are distributed without an appropriate treatment of convection and diffusion, especially for atmospheres that do not resemble Earth's.
%
%
%
%

The purpose of this paper is to introduce a newly developed three-dimensional non-hydrostatic model that simulates moist convection and cloud formation across various planetary conditions. Using a regional non-hydrostatic model to study moist convection is standard practice for Earth's atmosphere, and many numerical models -- named cloud-resolving models (CRM) or large eddy simulations (LES) -- have been developed accordingly \citep{Pielke92,Khairoutdinov03,Bryan02,Pressel15}. Similar non-hydrostatic models have been developed for Mars \citep{Rafkin01}, Jupiter \citep{Sugiyama14,Hueso01} and Saturn \citep{Hueso04}. But they lack a well-documented comparison against an accepted nonlinear solution, which seems to be the first step of developing a robust numerical model. Since there are no standard tests of non-hydrostatic models for planets other than Earth, we will primarily perform simulations against standard tests for Earth-like conditions. Then, we perform an idealized simulation of the Jovian atmosphere in radiative-convective equilibrium (RCE) to study moist convection in hydrogen atmospheres.

The structure of this work is organized as follows. Section \ref{sec:eom} is devoted to the equation of motion, which includes dry air, vapors, clouds, and precipitation. Section \ref{sec:mp} discusses the basic microphysics scheme employed in the model. Section \ref{sec:num} describes the numerical schemes that are designed for simulating atmospheric flows. Section \ref{sec:benchmark} provides a series of benchmark tests against known solutions in the literature. Section \ref{sec:jupiter} performs an idealized RCE simulation of the Jovian atmosphere. Section \ref{sec:conc} concludes and outlines the future applications. To make the notations consistent and clear, major symbols used in this paper are summarized in Appendix \ref{sec:symbols} for reference. 
%
%
%
%

\section{Equation of motion} \label{sec:eom}
We solve the most generic form of the equation of motion, the fully compressible Euler equations, to avoid unnecessary assumptions that are only valid for a particular type of planet. Though Euler equations can be written in many mathematically equivalent forms, they are different when discretized and implemented in a numerical model. \cite{Bryan02} has compared five different ways of implementing the Euler equations, using potential temperature, equivalent potential temperature, and liquid water potential temperature. They concluded that the form of governing equations used in a numerical model has a profound effect on the simulation of a warm rising bubble. \cite{Satoh02} pointed out that in a model with diabatic forcing, the change in the domain integral of total energy is generally different from energy gain/loss from the boundaries if the potential temperature is used as the prognostic variable, which causes an error of the energy budget that accumulates with time. Furthermore, the expressions of potential temperature and equivalent potential temperature become complicated when the heat capacity of the atmosphere varies with temperature or when multiple condensing species exist in the atmosphere. Given the limitations of using potential temperature, we use total energy as the prognostic variable in implementing the Euler equations. We build our hydrodynamic solver on top of the Athena++ framework, an efficient, scalable and well-tested astrophysical code \citep{White16,Stone08,Stone09}. Using the Athena++ framework as the software infrastructure not only saves us a tremendous amount of the development work but also enables the atmospheric part to utilize all the power of the Athena++ infrastructure such as the static/adaptive mesh-refinement, curvilinear geometry, and dynamic task scheduling. 
%
%
%
%

To simulate a heterogeneous atmosphere under the gravitational field, we extend the original Euler equations to vapors, clouds, and precipitation. The continuity equation for each homogeneous component should be solved individually, leading to a family of continuity equations. A heterogeneous air parcel comprises dry air ($\rho_d$), vapors ($\rho_i$), and condensates ($\rho_{ij}$), where $i$ is the index for the vapor and $j$ is the index for the phase of the condensate. A condensate is a general description of any condensed material, which may represent a liquid cloud, ice cloud, graupel, rain, snow, etc. Two indices $(i,j)$ are needed to identify a condensate because of the presence of multiple condensing species. Several simplifications to the true physical processes are made to limit the complexity of this model. We assume that the gaseous components share the same mean motion velocities ($u,v,w$) and the condensates have additional terminal velocities ($w^t_{ij}$) with respect to the mean motion of the air parcel. We treat each component in the atmosphere equally and solve the evolution of its density tendency separately as shown in equations (\ref{eqn:d1}-\ref{eqn:d3}). 
%
%
%
%
\begin{eqnarray}
\frac{\partial\rho_d}{\partial t}+\frac{\partial(\rho_d u)}{\partial x}+\frac{\partial(\rho_d v)}{\partial y}+\frac{\partial(\rho_d w)}{\partial z} & = & 0 \label{eqn:d1} \\
\frac{\partial\rho_i}{\partial t}+\frac{\partial(\rho_i u)}{\partial x}+\frac{\partial(\rho_i v)}{\partial y}+\frac{\partial(\rho_i w)}{\partial z} & = & 0 \label{eqn:d2} \\
\frac{\partial\rho_{ij}}{\partial t}+\frac{\partial(\rho_{ij} u)}{\partial x}+\frac{\partial(\rho_{ij} v)}{\partial y}+\frac{\partial(\rho_{ij} w)}{\partial z} & = & -\frac{\partial(\rho_{ij}w^t_{ij})}{\partial z} \label{eqn:d3}
\end{eqnarray}
Comparing to equations (\ref{eqn:d1}-\ref{eqn:d2}), the equation for condensates, equation (\ref{eqn:d3}), has an additional flux term $w^t_{ij} \rho_{ij}$ in the vertical direction due to the flux of sedimentation. Sedimentation of cloud droplets is found to be important for simulating stratocumulus cloud on Earth \citep{Ackerman04,Bretherton07} and probably for simulating the folded filament cloud on Jupiter as well. The value of terminal velocity $w^t_{ij}$ should be provided externally by a microphysics package specific to the application. Changes to the continuity equations due to thermodynamics or microphysics will be discussed separately in Section \ref{sec:mp}.
%
%
%
%

Precipitation and sedimentation not only transfer mass but transfer momentum and energy as well. The momentum equations now read:
\begin{eqnarray}
\frac{\partial (\rho u)}{\partial t}+\frac{\partial (u\rho u+p)}{\partial x}+\frac{\partial (v\rho u)}{\partial y}+\frac{\partial (w\rho u)}{\partial z} & = & -\sum_{i,j}\frac{\partial (w^t_{ij}\rho_{ij}u)}{\partial z} \\
\frac{\partial (\rho v)}{\partial t}+\frac{\partial (u\rho v)}{\partial x}+\frac{\partial (v\rho v+p)}{\partial y}+\frac{\partial (w\rho v)}{\partial z} & = & -\sum_{i,j}\frac{\partial (w^t_{ij}\rho_{ij}v)}{\partial z} \\
\frac{\partial (\rho w)}{\partial t}+\frac{\partial (u\rho w)}{\partial x}+\frac{\partial (v\rho w)}{\partial y}+\frac{\partial (w\rho w+p)}{\partial z} & = & -\rho g - \sum_{i,j}\frac{\partial (w^t_{ij}\rho_{ij}w)}{\partial z} \label{eqn:mom3},
\end{eqnarray}
where the total density $\rho$ is defined as:
\begin{equation}
\rho=\rho_d+\sum_i\rho_i+\sum_{i,j}\rho_{ij}
\end{equation}
Note that the total density includes the contribution from all gases and condensates regardless of whether sedimentation occurs or not. If the condensate is aloft in the air, its mass is part of the total mass of the air parcel. Otherwise, the frictional force during sedimentation equals the gravity of the condensate. Then, the term $\sum_{i,j}\rho_{ij}g$ can be interpreted as the drag force acted upon the air parcel by sedimentation.

The total energy of a heterogeneous air parcel is defined as:

\begin{equation}
\rho e = \rho_d c_{v,d}T + \sum_i\rho_i c_{v,i}T
	   + \sum_{i,j}\rho_{ij} c_{ij}T
	   + \frac{1}{2}\rho(u^2+v^2+w^2)
       + \sum_i\mu_i\rho_i
       + \sum_{i,j}\mu_{ij}\rho_{ij}, \label{eqn:energy}
\end{equation}
where the first three terms represent the internal energies, the fourth term is the kinetic energy, and the last two terms represent the chemical potentials. Because the chemical potential is defined only to within an additive constant, it is possible to set the chemical potentials of all gases ($\mu_i$) to zero so long as the gases are chemically independent. Otherwise, different chemically potentials should be assigned to reflect the differences in their internal energies. For example, in a hydrogen atmosphere, hydrogen in the ortho state (odd rotational quantum numbers, $J$) has higher internal energy than hydrogen in the para state (even $J$). The difference increases with decreasing temperature. Brought from the deeps, the ortho hydrogen slowly converts to the para hydrogen at a shallower depth on a time scale of about $10^8$ sec \citep{Conrath84}. The latent heat release associated with the conversion is thought to be important for the energy transport in hydrogen atmospheres \citep{Conrath98}. If such effect were to be calculated, ortho hydrogen and para hydrogen should be treated as different species and with different chemical potentials. 

The chemical potential of the condensate ($\mu_{ij}$) is related to the chemical potential of the corresponding gas ($\mu_i$) and the latent heat ($L_{ij}$). Using the Kirchhoff's equation\footnote{There are a lot of Kirchhoff's equations. What we mean is: $dL_{ij}(T)/dT=c_{p,i}-c_{ij}$}, the latent heat as a function of temperature is approximately \citep{Emanuel94}:

\begin{equation}
L_{ij}(T)=L^r_{ij} - \Delta c_{ij}(T-T^r) \label{eqn:latent},
\end{equation}
where $L^r_{ij}$ is latent heat of condensate $(i,j)$ at reference temperature $T^r$ and $\Delta c_{ij}=c_{ij}-c_{p,i}$ is the difference between the specific heat capacities. Latent heat is defined as enthalpy difference between the vapor and the condensate:

\begin{equation}
L_{ij}(T)=c_{p,i}T+\mu_i-(c_{ij}T+\mu_{ij})
\end{equation}
Therefore, the chemical potential of condensate $(i,j)$ is:

\begin{equation}
\mu_{ij} = \mu_i - (L_{ij}(T) + \Delta c_{ij} T) = \mu_i - (L^r_{ij}+\Delta c_{ij} T^r) \label{eqn:muij}
\end{equation}

Assuming that gas $k$ undergoes phase transition, in which the mass transfer is $\Delta \rho$ and the condensate is $(k,l)$, the equation of the conservation of energy is: 
%
%
%
%

\begin{equation}
(\rho_d c_{v,d} + \sum_i\rho_i c_{v,i}
	   + \sum_{i,j}\rho_{ij} c_{ij})T =
(\rho_d c_{v,d} + \sum_i\rho_i c_{v,i}
	   + \sum_{i,j}\rho_{ij} c_{ij}-\Delta\rho c_{v,k}+\Delta\rho c_{kl})(T + \Delta T)
       + (\mu_{kl}- \mu_k)\Delta\rho \label{eqn:phase},
\end{equation}
where $\Delta T$ is the temperature change. Using equations (\ref{eqn:latent}), (\ref{eqn:muij}), and (\ref{eqn:phase}) to solve for $\Delta T$ yields:
%
%
%
%

\begin{equation}
\Delta T=\frac{\Delta\rho(L_{kl}-R_kT)}{\rho_d c_{v,d} + \sum_i\rho_i c_{v,i}+\sum_{i,j}\rho_{ij} c_{ij}-\Delta\rho c_{v,k}+\Delta\rho c_{kl}}
\end{equation}
The numerator is the latent heat release, and the denominator is the heat capacity of the air parcel after condensation. Note that the latent heat release at constant volume is different from the latent heat release at constant pressure, which is typically defined and used in a large-scale GCM. 

The equation of total energy for atmospheric flows has forcing terms from gravity, friction, and sedimentation, which is:

\begin{equation}
\frac{\partial(\rho e)}{\partial t}+\frac{\partial[u(\rho e+p)]}{\partial x}+\frac{\partial[v(\rho e+p)]}{\partial y}+\frac{\partial[w(\rho e+p)]}{\partial z} = -\rho gw-\sum_{i,j}\rho_{ij}gw^t_{ij}-\sum_{i,j}\frac{\partial (w^t_{ij}\rho_{ij}e_{ij})}{\partial z}, 
\end{equation}
where $e_{ij}=c_{ij}T+\frac{1}{2}(u^2+v^2+w^2)+\mu_{ij}$ is the specific total energy of condensate $(i,j)$. The first term on the right-hand-side is the forcing from gravity, the second term is the frictional heating from sedimentation, and the third term is the divergence of the energy flux due to sedimentation. 

Finally, the Euler equations are closed by the equation of state, which is simply $p=\rho R_d T$ for a homogeneous ideal gas. Yet, in the presence of gases and clouds, the equation of state is obtained by summing the partial pressure of all gaseous components. Let $\epsilon_i$ be the molecular weight ratio of the vapor to the dry air. The total pressure is:

\begin{equation}
\begin{aligned}
p & = \rho_d R_d T + \sum_i \rho_i R_i T \\
  & = \rho R_d T(q_d+\sum_i q_i/\epsilon_i) \\
  & = \rho R_d T\big[1+\sum_i q_i(1/\epsilon_i-1)-\sum_{i,j} q_{ij}\big] \\
  & = \rho R_d T_v \,, \label{eqn:eos}
\end{aligned}
\end{equation}
where the virtual temperature $T_v$ is define as:

\begin{equation}
T_v = T\big[1+\sum_i q_i(1/\epsilon_i-1)-\sum_{ij} q_{ij}\big] \label{eqn:tv}
\end{equation}
In an atmosphere laden with vapors and clouds, it is the virtual temperature rather than the actual temperature that determines the density of an air parcel. Thus, a moist air parcel loaded with water vapor in Jupiter's atmosphere $(\epsilon_i>1)$ is denser than a dry air parcel at the same temperature and pressure.
%
%
%
%

%
%
%
%

\section{Microphysics and hydrological cycle} \label{sec:mp}
Simulating the formation and dissipation of clouds is indispensable in a non-hydrostatic and cloud-resolving model. In the presence of abundant condensation nuclei, cloud forms when the partial pressure of the vapor exceeds its saturation vapor pressure. Then cloud particles undergo coagulation and coalescence to grow in size. When they become large enough to have appreciable falling velocities, precipitation occurs. The falling rain droplets or snowflakes re-evaporate into unsaturated warm air below the cloud, adding mass to the air parcel and reducing its temperature. The whole subject of microphysical processes regarding clouds is very complicated and is still not very well understood especially for how cloud forms on other planets. Moreover, most microphysical packages are developed specifically for water clouds on Earth (e.g. \citealt{Kessler69,Lin83,Seifert01}). How to adapt the existing parameterizations developed for Earth to planets is a grand topic open for future study (some advances can be found at \citealt{Rossow78,Carlson88}). It should be made clear that the subject of this section is not to compare and contrast different microphysics schemes, nor to design a new one, but rather, to describe a unified framework of implementing microphysics in a dynamic model, that works for both terrestrial planets and giant planets. It may seem unnecessary at this point for the dynamic part (core) of a numerical model is usually separated from other physical packages in a traditional GCM. However, since we are using total energy (including chemical potentials) as the prognostic variable of the thermodynamic equation, the microphysics is so intimately related to dynamics that should be implemented in a way that is congruent with the formulation of the dynamics. 
%
%
%
%

The fact that dynamic model stays with density of the dry air ($\rho_d$) and total energy ($\rho e$) as prognostic variables eases the implementation of the microphysics because those two fields are also conserved during phase change. What remains is to calculate the partition of the vapors ($\rho_i$) and the corresponding condensates ($\rho_{ij}$) subject to the specific formulation of the microphysics. Assuming that the reaction rates are described by abstract functions:

\begin{eqnarray}
\frac{\partial q_i}{\partial t} & = & \Phi_i(\rho,q_i,q_{ij},u,v,w,p) \label{eqn:qi1} \\
\frac{\partial q_{ij}}{\partial t} & = & \Phi_{ij}(\rho,q_i,q_{ij},u,v,w,p), \label{eqn:qij}
\end{eqnarray}
where $q_i=\rho_i/\rho$ and $q_{ij}=\rho_{ij}/\rho$ are mass mixing ratios. The reaction rates are written as functions of primitive variables $\mathbf{X}=(\rho,q_i,q_{ij},u,v,w,p)^T$ since microphysics or chemical reactions usually works with mixing ratios rather than densities. The conversion between the conserved variables $\mathbf{Y}=(\rho_d,\rho_i,\rho_{ij},\rho u,\rho v,\rho w, \rho e)^T$ and the primitive variables $\mathbf{X}$ is detailed in Appendix \ref{sec:prim}. 

The evolution equations (\ref{eqn:qi1}) and (\ref{eqn:qij}) can be integrated using a standard ODE integrator, such as the first-order backward Euler integration method:

\begin{equation}
\frac{\mathbf{X}^{n+1,m}-\mathbf{X}^n}{\Delta t}=\mathbf{\Phi}(\mathbf{X}^{n+1,m})\approx
\mathbf{\Phi}(\mathbf{X}^{n+1,m-1})+\frac{\partial\mathbf{\Phi}}{\partial \mathbf{X}}(\mathbf{X}^{n+1,m}-\mathbf{X}^{n+1,m-1}),
\end{equation}
where $\frac{\partial\mathbf{\Phi}}{\partial \mathbf{X}}$ is the Jacobian matrix and $m$ is the iteration number. Iteration is needed because a linear expansion is used to approximate $\mathbf{\Phi}(\mathbf{X}^{n+1})$. The iteration starts with:

\begin{equation}
\mathbf{X}^{n+1,0}=\mathbf{X}^n,
\end{equation}
and progresses forward with:

\begin{equation}
\mathbf{X}^{n+1,m}=\mathbf{X}^{n+1,m-1}+(\frac{\mathbf{I}}{\Delta t}-\frac{\partial\mathbf{\Phi}}{\partial \mathbf{X}})^{-1}\big[\mathbf{\Phi}(\mathbf{X}^{n+1,m-1})-\frac{\mathbf{X}^{n+1,m-1}-\mathbf{X}^n}{\Delta t}\big]
\label{eqn:x_n+1}
\end{equation}
The iteration stops when $||\mathbf{X}^{n+1,m}-\mathbf{X}^{n+1,m-1}||$ is smaller than an arbitrary small number. Then, at the last iteration:

\begin{eqnarray}
\mathbf{X}^{n+1,m} & = & \mathbf{X}^{n,m-1} = \mathbf{X}^{n+1} \\
\frac{\mathbf{X}^{n+1}-\mathbf{X}^n}{\Delta t} & = & \mathbf{\Phi}(\mathbf{X}^{n+1}),
\end{eqnarray}
which completes the backward Euler integration method.

Equation (\ref{eqn:x_n+1}) requires both the reaction rate $\mathbf{\Phi}(\mathbf{X})$ and the Jacobian matrix of the reaction rate $\frac{\partial\mathbf{\Phi}}{\partial \mathbf{X}}$ to be evaluated at each time step. Although equation (\ref{eqn:x_n+1}) is only first-order accurate in time, it is unconditionally monotone and is especially good at integrating stiff equations in which the chemical timescale (such as condensation) is much shorter than the dynamic time step. A second-order accurate and L-stable integration method like TR-BDF2 \citep{Hosea96} can also be used to integrate equations (\ref{eqn:qi1}) and (\ref{eqn:qij}). However, given the uncertainties in the reaction rate coefficients, using the backward Euler scheme is probably enough for simulating planetary atmospheres. The above procedure costs more in computational time than the simple Euler forward method, but it is more robust regarding numerical stability and can be generically applied to any chemical systems. 

Here we give an example of how the Kessler scheme \citep{Kessler69} is implemented according to the above framework. The Kessler scheme is the simplest cloud microphysics scheme that represents the cycling of the three phases of a condensible gas (vapor, cloud, and precipitation). Although it is considered outdated for modeling Earth's atmosphere, it is still of great value when applied to planetary atmospheres due to its simplicity. In fact, many cloud-resolving simulations for the Jovian atmosphere, including \cite{Yair95,Nakajima00,Hueso04,Sugiyama14}, started with microphysics schemes similar to \cite{Kessler69}. The original paper of \cite{Kessler69} was specifically written for Earth's atmosphere, with hardwired physical constants in the equations. We rewrite the formula and summarize the major equations in the following:

\begin{eqnarray}
\frac{\partial q_i}{\partial t} & = & -k_1(q_i-q^*_{i1})^+ + k_4q_{i2}(q^*_{i1}-q_i)^+ \label{eqn:qi} \\ 
\frac{\partial q_{i1}}{\partial t} & = & k_1(q_i-q^*_{i1})^+ - k_3q_{i1}q_{i2} - k_2q_{i1} \\
\frac{\partial q_{i2}}{\partial t} & = & -k_4q_{i2}(q^*_{i1}-q_i)^+ + k_3q_{i1}q_{i2} + k_2q_{i1} \label{eqn:qi2} \\ 
w^t_{i1} & = & 0 \\
w^t_{i2} & = & k_5,
\end{eqnarray}
where $k_1,k_2,k_3$, and $k_4$ are reaction rate coefficients concerning the rate of condensation, the rate of autoconvection (cloud becomes precipitation), the rate of accretion (precipitation grows bigger by collecting cloud particles), and the rate of evaporation, respectively. $k_5$ is a physical constant describing the terminal velocity of the precipitation. Sedimentation of cloud particles was not considered in \cite{Kessler69}. $q_{i1}$ is the mass mixing ratio of the cloud, $q_{i2}$ is the mass mixing ratio of the precipitation, and $q^*_{i1}$ is the saturation vapor mixing ratio with respect to cloud $q_{i1}$. The symbol $x^+=max(x,0)$. 

\begin{figure}[ht!]
\centering
\includegraphics[scale=0.5]{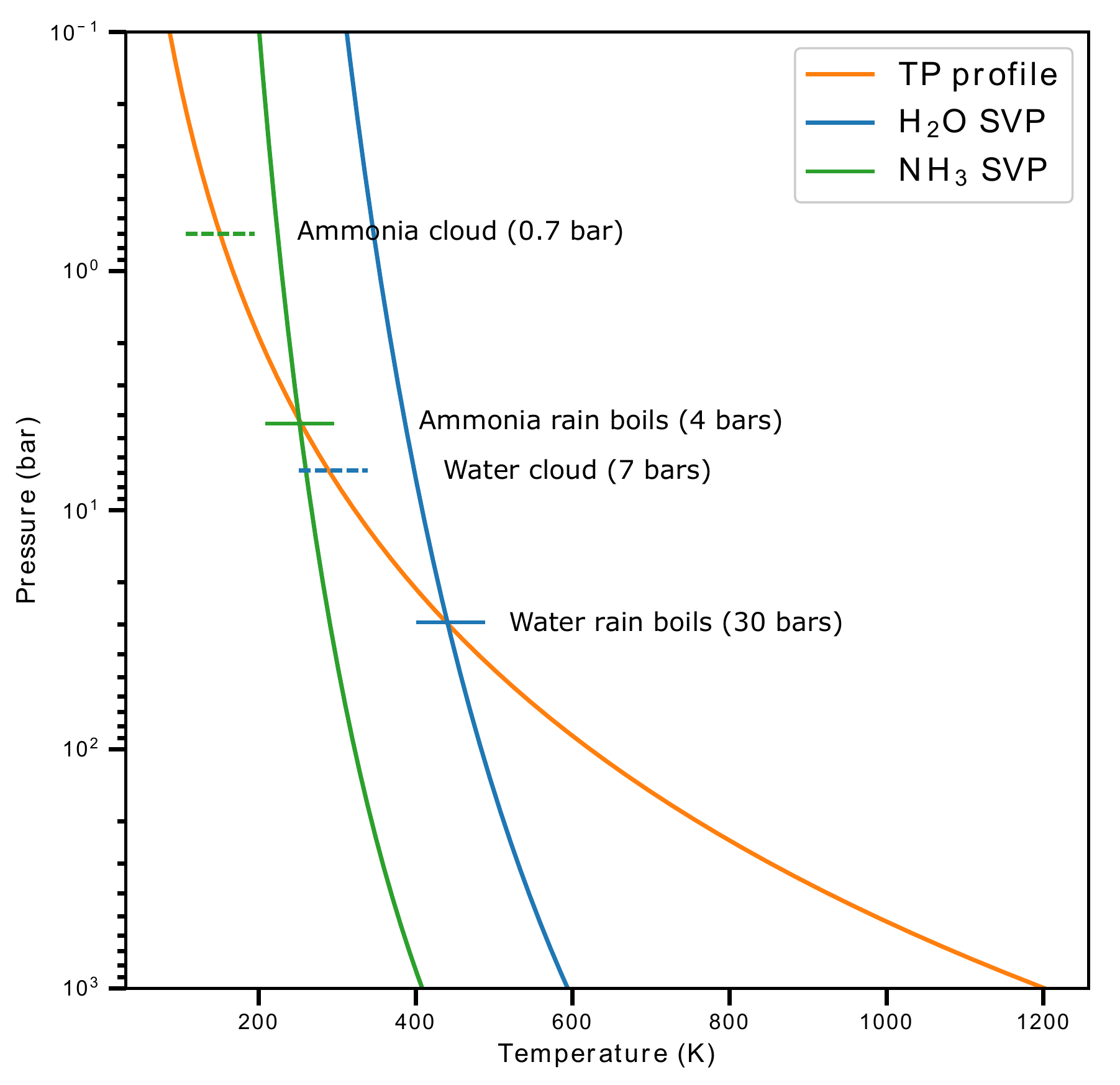}
\caption{Boiling levels for water and ammonia rain droplets in Jupiter's atmosphere. Blue line is the saturation vapor pressure of liquid water. Green line is the saturation vapor pressure of liquid ammonia. Orange line is the temperature profile of Jupiter's atmosphere.}
\label{fig:svp_tp}
\end{figure}

Coefficient $k_1$ controls how fast supersaturated vapor condenses to form clouds. If the condensation nuclei are abundant, condensation can occur almost instantaneously when the vapor is saturated. Otherwise, spontaneous nucleation requires a high degree of supersaturation. A larger value of $k_1$ produces less supersaturation. Coefficient $k_2$ controls how fast clouds start to precipitate, and thus the amount of cloud present in the atmosphere. Coefficient $k_3$ functions similarly as $k_2$, but the accretion rate $k_3q_{i1}q_{i2}$ is a quadratic term depending both on the amount of cloud and the amount of precipitation. Coefficient $k_4$ controls how fast precipitation evaporates. The evaporation rate is proportional to the amount of precipitation $q_{i2}$ and the amount of saturation deficit $(q^*_{i1}-q_i)^+$. The values of these coefficients depend largely on particle number density, which can be either empirically prescribed or physically calculated via a more sophisticated two-moment microphysics scheme such as \cite{Seifert06}. Often, the exact values of these coefficients have little impact on the dynamics as long as they are chosen within a reasonable range, which can be informed by Earth's conditions. \cite{Sugiyama14} has varied the autoconversion rate ($k_2$) for two orders of magnitude, and they concluded that intermittent outburst of convective activity found in their simulation did not change greatly with $k_2$. In addition to the microphysical reactions outlined in equations (\ref{eqn:qi}) -- (\ref{eqn:qi2}), precipitation ($q_{i2}$) may \textit{boil} in the atmosphere when the saturation vapor pressure equals the ambient atmospheric pressure. Boiling of rain droplets is not likely to occur in the terrestrial environment but must happen in Jovian atmospheres if the evaporation rate is low (large droplet size), and if the precipitation survives evaporation before reaching the boiling level -- 30 bars for water and 4 bars for ammonia in the current Jupiter's atmosphere (shown in Figure \ref{fig:svp_tp}). The dynamic effect of boiling rain droplets is also an interesting topic worth future investigation. 

The Jacobian matrix of the reaction rate has two expressions depending on whether the atmosphere is saturated or not. For a saturated atmosphere ($q_i>q^*_{i1}$):

\begin{equation}
\frac{\partial\mathbf{\Phi}}{\partial \mathbf{X}}=
\begin{pmatrix} 
-k_1 & 0 & 0 \\ 
k1 & -k_2-k_3q_{i2} & -k_3q_{i1} \\
0 & k_2+k_3q_{i2} & k_3q_{i1}
\end{pmatrix}
\end{equation}
For an unsaturated atmosphere ($q_i<q^*_{i1}$):

\begin{equation}
\frac{\partial\mathbf{\Phi}}{\partial \mathbf{X}}=
\begin{pmatrix} 
-k_4q_{i2} & 0 & 0 \\ 
0 & -k_2-k_3q_{i2} & -k_3q_{i1} \\
k_4q_{i2} & k_2+k_3q_{i2} & k_3q_{i1}
\end{pmatrix}
\end{equation}
Since the Jacobian matrix is only $3\times 3$ in this case, using the backward Euler integration method as described by equation (\ref{eqn:x_n+1}) is almost as fast as the simple Euler forward method with significant improvement in the numerical stability.

\section{Numerical schemes} \label{sec:num}
The hydrodynamic equations are discretized using the finite-volume method, in which all conserved quantities are placed in a collocated grid, representing the spatial average of the finite volume. Let $Q_{i,j,k}^n$ be any discretized conserved quantity at the cell $(i,j,k)$ and the $n$-th time. The simplest first-order accurate evolution equation in time is:

\begin{equation}
\begin{aligned}
Q_{i,j,k}^{n+1} = Q_{i,j,k}^n \label{eqn:forward}
			& - \frac{\Delta t}{\Delta x}(F_{i+1/2,j,k}^n-F_{i-1/2,j,k}^n) \\
 			& - \frac{\Delta t}{\Delta y}(G_{i,j+1/2,k}^n-G_{i,j-1/2,k}^n) \\
 			& - \frac{\Delta t}{\Delta z}(H_{i,j,k+1/2}^n-H_{i,j,k-1/2}^n)+\Delta  t\mathscr{F}_{i,j,k}^n, 
\end{aligned}
\end{equation}
where $F_{i-1/2,j,k}^n$ is the numerical flux between cells $(i-1,j,k)$ and $(i,j,k)$ at the $n$-th time step, and $\mathscr{F}_{i,j,k}$ is the body force in the right-hand-side of the hydrodynamic equations. The meanings of $G$ and $H$ are similar, but denote the fluxes in the $y-$ and $z-$ directions. Since the fluxes in all directions are calculated by the Riemann solver, we will focus on the $x$-direction and omit the $j,k$ indices in the subscript and the $n$ in the superscript for clarity. 
%
%
%
%

The numerical flux $F_{i-1/2}$ is the solution of an initial value problem, the Riemann problem $\mathcal{R}$, consisting two states that are at the left ($Q_{i-1/2}^{-}$) and the right ($Q_{i-1/2}^{+}$) side of a jump discontinuity.
\begin{equation}
F_{i-1/2}=\mathcal{R}(Q_{i-1/2}^{-},Q_{i-1/2}^{+}) \label{eqn:riemann}
\end{equation}
The left and right states are interpolated from cell-averaged quantities, a process called subgrid-reconstruction. A $(2m-1)$-th order reconstruction scheme takes the form of:
\begin{eqnarray}
Q_{i-1/2}^{-} & = & \mathcal{I}(Q_{i-m},Q_{i-m+1},\ldots,Q_{i-1},Q_i,\ldots,Q_{i+m-2}) \\
Q_{i-1/2}^{+} & = & \mathcal{I}(Q_{i+m-1},Q_{i+m-2},\ldots,Q_i,Q_{i-1},\ldots,Q_{i-m+1}), \label{eqn:q+}
\end{eqnarray}
where $\mathcal{I}$ is the interpolation function and is biased toward one side of the discontinuity. Equations (\ref{eqn:forward}) - (\ref{eqn:q+}) summarized the standard Godunov method of solving hyperbolic equations \citep{LeVeque02}, which were implemented in the hydrodynamic solver of the Athena++ code (\citet{Stone08,Stone09}). However, because the atmospheric flows usually are at low Mach number and are stratified in the vertical direction, the solver designed for supersonic astrophysical flows should be modified according to the characteristics of the atmospheric flow. We elaborate our extension of the numerical methods of Riemann solver, subgrid-reconstruction, and time-stepping to atmospheric flows in the following subsections.

\subsection{Riemann solver} \label{sec:riemann}
The standard Riemann solvers in the Athena++ model are Roe solver \citep{Roe81} and HLLC solver \citep{Toro13}. Applying the default solvers to atmospheric problems will induce spurious errors because the gravitational acceleration nearly balances the vertical pressure gradient. The small imbalance is what drives the vertical motion, but would be obscured by the interpolation error if the stratification in pressure is not taken into account in the reconstruction step or the Riemann solver (demonstrated in Figure \ref{fig:straka1} and Figure \ref{fig:straka2}). Many remedies have been proposed to alleviate the numerical difficulty in the quasi-steady problems including subtracting a stationary state \citep{Dedner01}, recognizing the pressure difference in the Riemann solver \citep{Bale03}, incorporating the hydrostatic equilibrium in the pressure reconstruction \citep{Kaeppeli16}, etc. Here we use the recently developed Low Mach number Approximate Riemann Solver (LMARS) \citep{Chen13} to solve equation (\ref{eqn:riemann}).

The LAMRS solver has several advantages over other approaches for atmospheric problems. First, as the Mach number approaches zero, the compressible Euler equations converge to the incompressible limit. Yet, the standard Riemann solver that captures shock, such as the Roe solver, failed to produce the convergence \citep{Guillard99}. The LAMRS solver is designed for low Mach number problems but can also handle the case in which the flow speed approaches the sound speed, bring an ideal solution to atmospheric flows. Second, the formulation of LMARS allows a seamless transition from hydrostatic models to non-hydrostatic models, a feature that will be useful for an adaptive mesh refined model. Third, the LAMRS scheme is easier to implement and faster to calculate than the Roe or HLLC solver. Its key results are:
\begin{eqnarray}
p_{i-1/2} & = & \frac{1}{2}(p_{i-1/2}^- +p_{i-1/2}^+)-\frac{\rho c}{2}(u_{i-1/2}^+ - u_{i-1/2}^-) \label{eqn:rp} \\
u_{i-1/2} & = & \frac{1}{2}(u_{i-1/2}^- +u_{i-1/2}^+)-\frac{1}{2\rho c}(p_{i-1/2}^+ - p_{i-1/2}^-) \label{eqn:ru},
\end{eqnarray}
where $p^\pm_{i\pm1/2}$ denotes the reconstructed pressure in the horizontal direction or the reconstructed perturbation pressure in the vertical direction. Knowing the pressure $p_{i-1/2}$ and the velocity $u_{i-1/2}$ at the cell boundary, it is easy to calculate the flux $F_{i-1/2}$. More details of the scheme are presented in \citet{Chen13} and not repeated here.

\subsection{subgrid-reconstruction}
The Athena++ model has implemented two reconstruction methods, the piecewise-linear method (PLM) and the piecewise-parabolic method (PPM, \cite{Colella84}). The PLM model is only second-order accurate in space, which would cause large phase error in the wave propagation when collocated variables are used \citep{Chen18}. The PPM model has a higher order of accuracy but does not have an asymmetry between the left and right side of the cell boundary, which is required in the LMARS scheme. Because of the aforementioned shortcomings, we implement the fifth-order Weighted Essentially Non-Oscillatory (WENO) scheme \citep{Shu98} in lieu of the original ones. The fifth-order WENO scheme is computed as convex combinations of three third-order stencils:

\begin{equation}
\begin{aligned}
\mathcal{I}_{weno}(Q_{i-3},Q_{i-2},Q_{i-1},Q_i,Q_{i+1}) \label{eqn:weno5}
	& = \alpha_0(\frac{1}{3}Q_{i-3}-\frac{7}{6}Q_{i-2}+\frac{11}{6}Q_{i-1}) \\
    & + \alpha_1(-\frac{1}{6}Q_{i-2}+\frac{5}{6}Q_{i-1}+\frac{1}{3}Q_i) \\
    & + \alpha_2(\frac{1}{3}Q_{i-1}+\frac{5}{6}Q_i-\frac{1}{6}Q_{i+1}),
\end{aligned}
\end{equation}
The weights $\alpha_0, \alpha_1$ and $\alpha_2$ are nonnegative and add up to one; they control which stencil to use. For a smooth field, $\alpha_0=0.1, \alpha_1=0.6, \alpha_2=0.3$, and equation (\ref{eqn:weno5}) reduces to the fifth-order polynomial interpolation scheme:

\begin{equation}
\mathcal{I}_{poly}(Q_{i-3},Q_{i-2},Q_{i-1},Q_i,Q_{i+1})=
\frac{1}{30}Q_{i-3}-\frac{13}{60}Q_{i-2}+\frac{47}{60}Q_{i-1}+\frac{9}{20}Q_i-\frac{1}{20}Q_{i+1}
\end{equation}

If a discontinuity exists, the weights are adjusted such that the stencil is placed toward the direction away from the discontinuity. The formulas for the weights are provided in \cite{Jiang96} and \cite{Shu98}. Following the convention of the original Athena++ model, the subgrid-reconstruction are performed using primitive variables ($\mathbf{X}$) instead of the conserved variables ($\mathbf{Y}$).

\subsection{time-stepping}
Equation (\ref{eqn:forward}) is only first-order accurate in time. To achieve higher order accuracy, a multi-stage time stepping method shall be used. A widely used third-order total variation diminishing (TVD) Runge-Kutta method is \citep{Shu88}:
\begin{eqnarray}
Q_i^\dagger & = & Q_i^n-\frac{\Delta t}{\Delta x}(F_{i+1/2}^n-F_{i-1/2}^n)+\Delta t\mathscr{F}_i^n \\
Q_i^\ddagger & = & \frac{3}{4}Q_i^n + \frac{1}{4}[Q_i^\dagger-\frac{\Delta t}{\Delta x}(F_{i+1/2}^\dagger-F_{i-1/2}^\dagger)+\Delta t\mathscr{F}_i^\dagger] \\
Q_i^{n+1} & = & \frac{1}{3}Q_i^n + \frac{2}{3}[Q_i^\ddagger-\frac{\Delta t}{\Delta x}(F_{i+1/2}^\ddagger-F_{i-1/2}^\ddagger)+\Delta t\mathscr{F}_i^\ddagger],
\end{eqnarray}
where $F_{i-1/2}^\dagger$, $F_{i-1/2}^\ddagger$ and $\mathscr{F}_i^\dagger$, $\mathscr{F}_i^\ddagger$ are fluxes and body forces evaluated using intermediate states $Q_i^\dagger$ and $Q_i^\ddagger$. By default, the above third-order scheme is used. Using the original second-order schemes such as the van Leer integrator is also possible.

\section{Benchmark tests} \label{sec:benchmark}
To illustrate that our model formulation and numerical scheme are suitable for atmospheric simulation, we perform three benchmark tests. The first one, proposed by \citet{straka93}, simulates a dense sinking bubble and its subsequent propagation on the lower boundary. This case is targeted to validate how well the numerical model handles a solid lower boundary. The second case, proposed by \citet{Robert93}, simulates a buoyant rising bubble and the fully developed Kelvin-Helmholtz instability. The forcing is weak in this case so that the turbulence field is sensitive to the numerical scheme. The third case, proposed by \citet{Bryan02}, is similar to the second one but includes moisture and phase change. This case is intended for testing the interaction between the dynamics and the thermodynamics.  

\subsection{Straka density current}\label{sec:straka}
A standard test of a non-hydrostatic model is to simulate a sinking bubble and the propagation of the resulting density current. It has been widely used as a test case for many numerical models (e.g. \cite{Skamarock93}, \cite{Ooyama01}, \cite{Pressel15}) since \citet{straka93} formulated this classic problem. The background atmosphere has a constant potential temperature of 300 K referenced at 1 bar. A cold bubble is put aloft in the air with maximum temperature difference -15 K. Then the bubble sinks and propagates after it hits the ground. 
%
%
%
%
%
%
%
%
%
%

\begin{figure}[ht!]
\plottwo{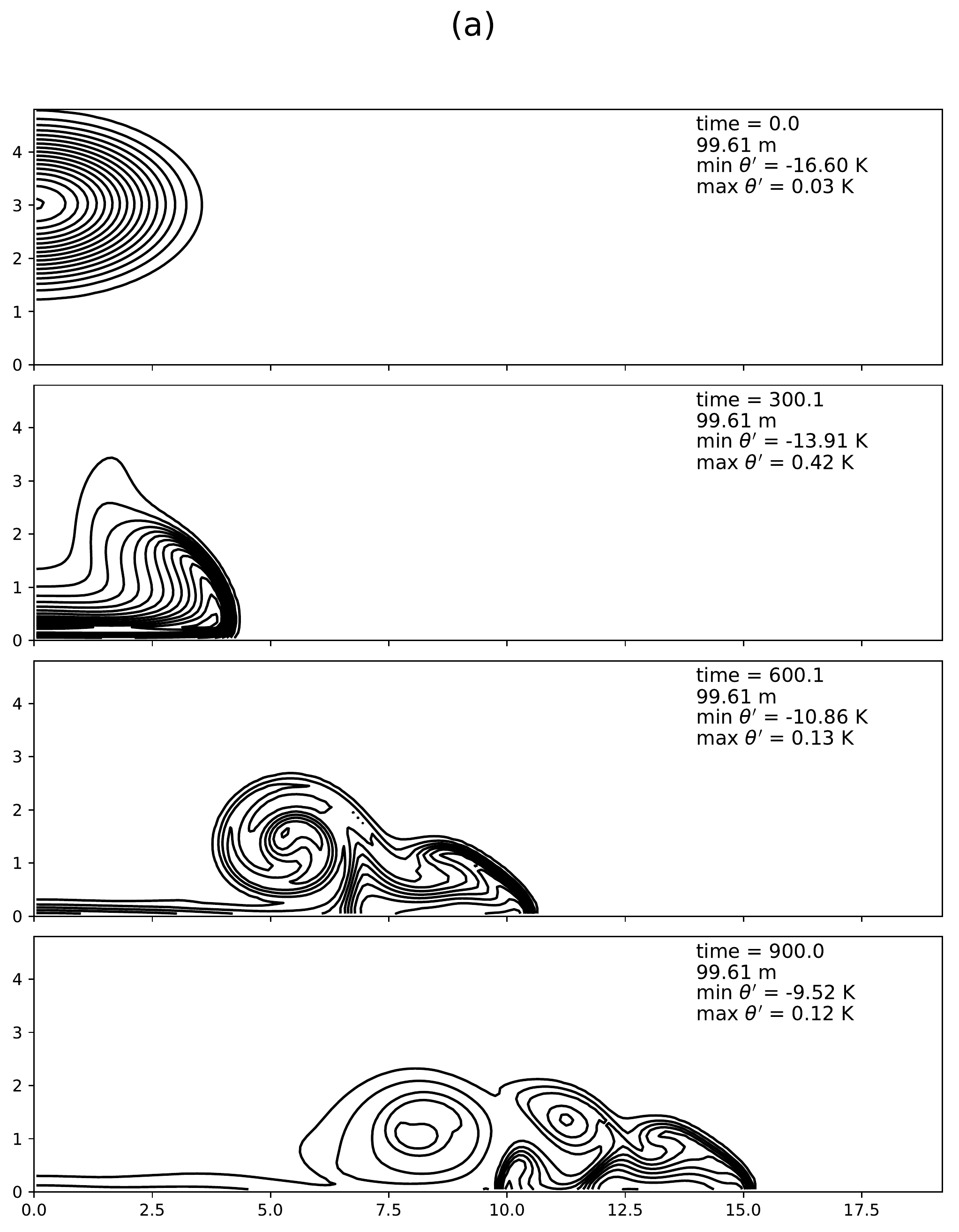}{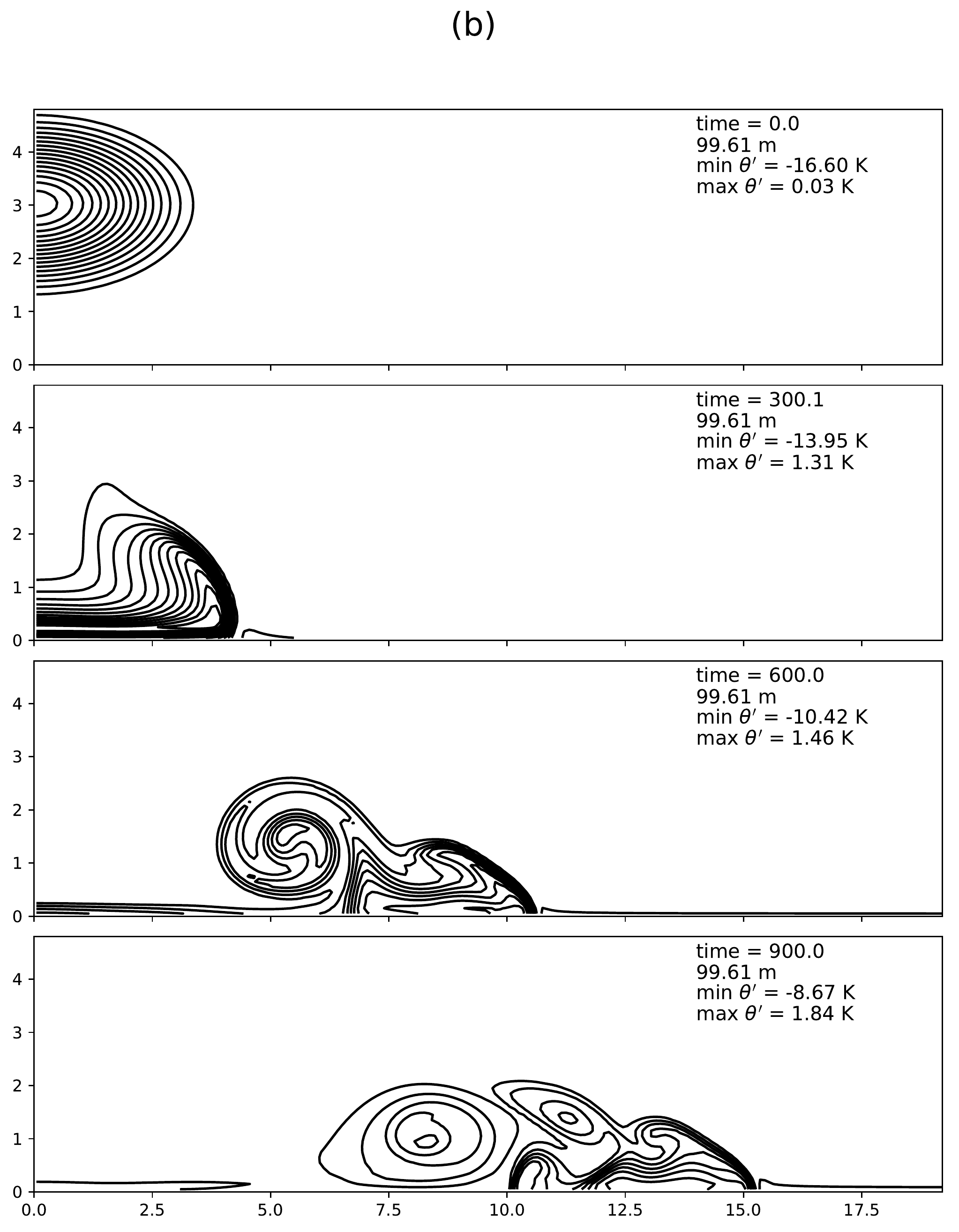}
\caption{Evolution of a dense sinking bubble at 0, 300, 600 and 900 seconds. Contours show the potential temperature anomaly. The contour interval is 1 K starting at -1 K. The time, resolution, minimum and maximum potential anomaly are indicated in each panel. a) the viscous solution as described by \cite{straka93} using LMARS solver. b) same solution using HLLC solver.}
\label{fig:straka1}
\end{figure}

The original setup specified by \citet{straka93} applied viscous dissipations to both the momentum equations and the energy equation to ensure the convergence of the solution. Here, we test the viscous solution for the comparison purpose as well as the nearly inviscid solution, in which no explicit diffusion is applied, to demonstrate the stability and the robustness of the numerical scheme. Figure \ref{fig:straka1}(a) shows the viscous solution using LMARS solver, which is almost identical to the reference solution provided by \citet{straka93}'s Figure 1. However, if the HLLC solver was used, the potential temperature at the surface has unrealistically increased by over 1 K ahead of the density current (Figure \ref{fig:straka1}b). The maximum potential temperature anomaly at 900s is 1.84 K using HLLC solver, while it is only 0.12 K using LMARS solver. In an adiabatic atmosphere, the potential temperature is very close to a conserved quantity. Therefore, in the inviscid limit, the maximum potential temperature anomaly should be zero, and the minimum potential temperature anomaly should be -15 K, the same as the initial condition. 
%
%

\begin{figure}[ht!]
\plottwo{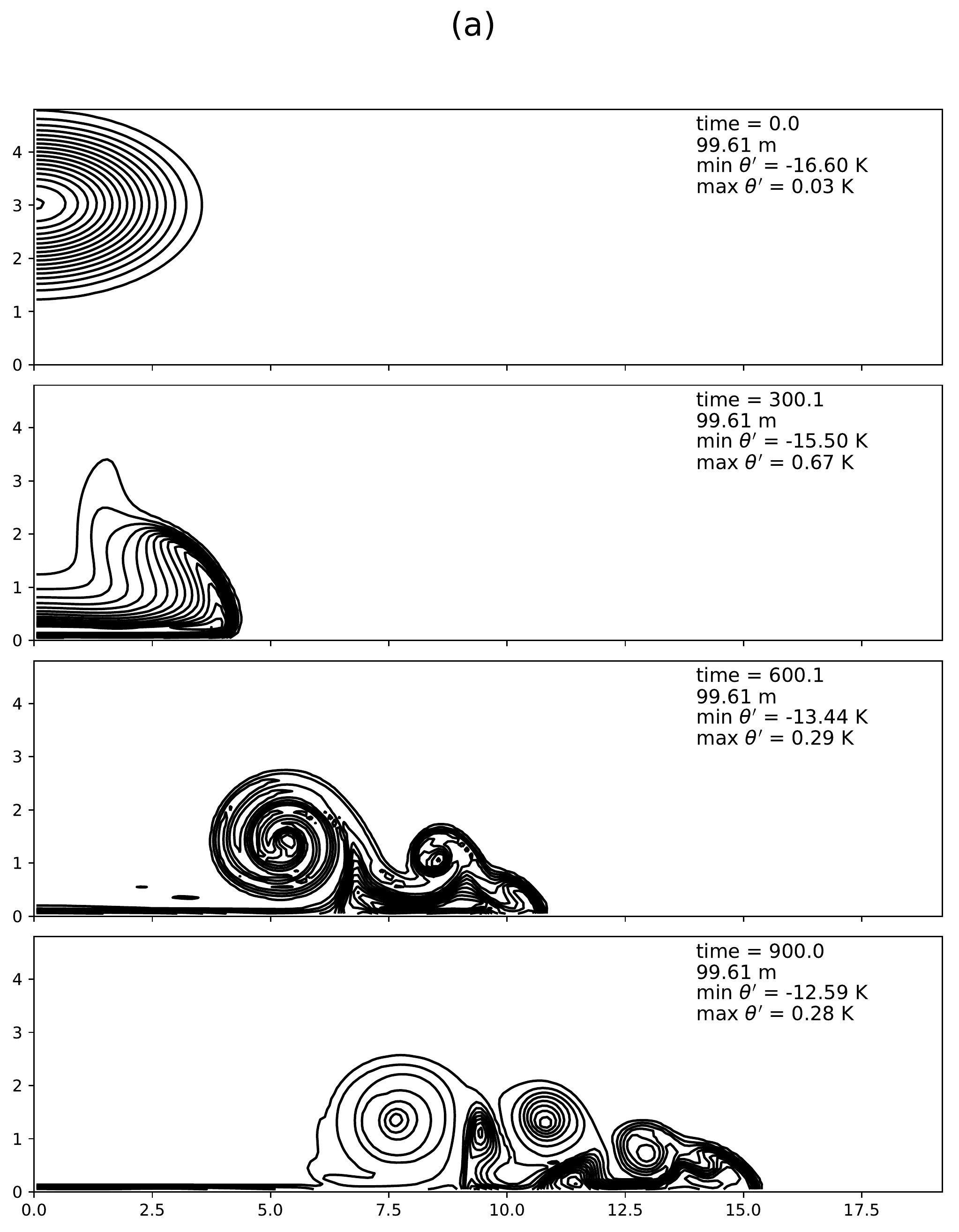}{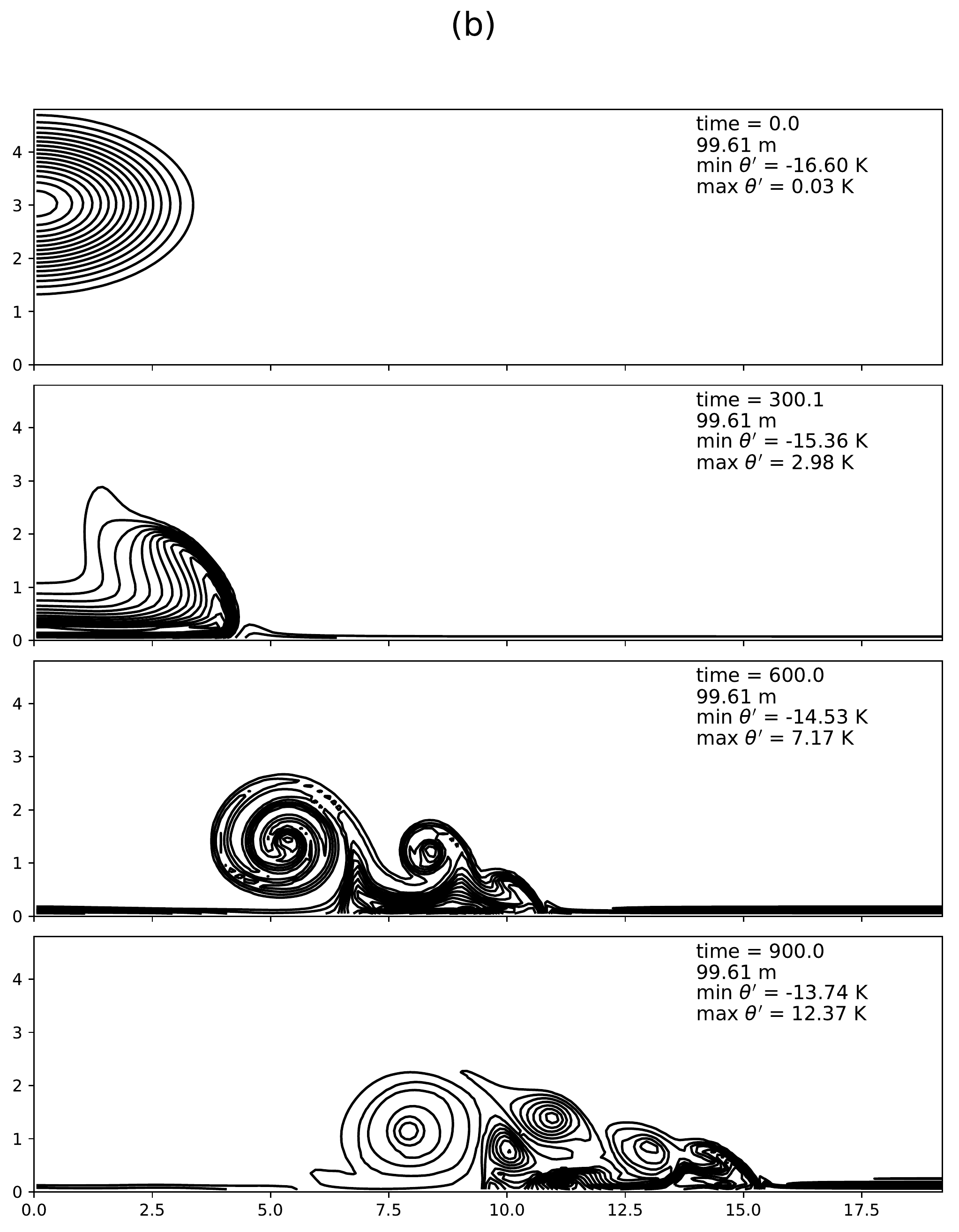}
\caption{Same as Figure \ref{fig:straka1} but for nearly inviscid solution. a) LMARS solver. b) HLLC solver}
\label{fig:straka2}
\end{figure}

As pointed out in \citet{Ooyama01}, it is possible to run the same simulation without any explicit diffusion terms, which results in the inviscid solution. Of course, a truly inviscid solution cannot be reached because numerical viscosity always exists as long as the mesh size is finite. The adjective ``inviscid'' refers to the formulation without explicit diffusion terms. The inviscid model is useful to test the stability of the numerical scheme. Figure \ref{fig:straka2} shows the inviscid solution, which generates more rotary eddies and sharper potential temperature gradient than the viscous one. The boundary condition becomes more problematic when explicit diffusion is disabled, and HLLC solver is used. HLLC solver produces unrealistic potential temperature perturbation of more than 12 K at 900s near the surface (Figure \ref{fig:straka2}b), while potential temperature perturbation remains as low as 0.3 K using LMARS solver (Figure \ref{fig:straka2}a). The solution we obtained using the LMARS solver at 600s is similar to \citet{Ooyama01}'s result in their Figure C3 and \citet{Guerra16}'s result in their Figure 12, confirming the veracity of the model. 

\subsection{Robert rising bubble}
\begin{figure}[ht!]
\plotone{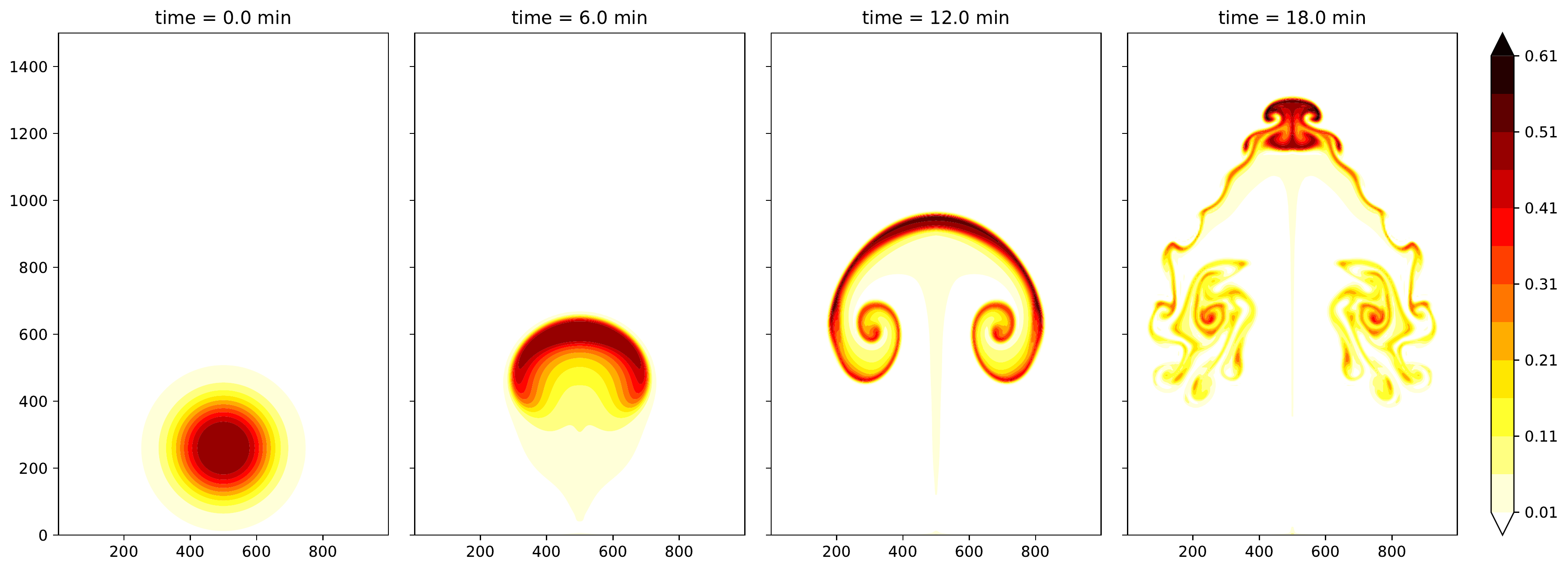}
\caption{Potential temperature anomalies of a Gaussian-shaped rising bubble. Time is indicated in the title of each panel. Spatial resolution is 5m. Domain size is 1km by 1.5km.}
\label{fig:robert}
\end{figure}

\begin{figure}[ht!]
\plotone{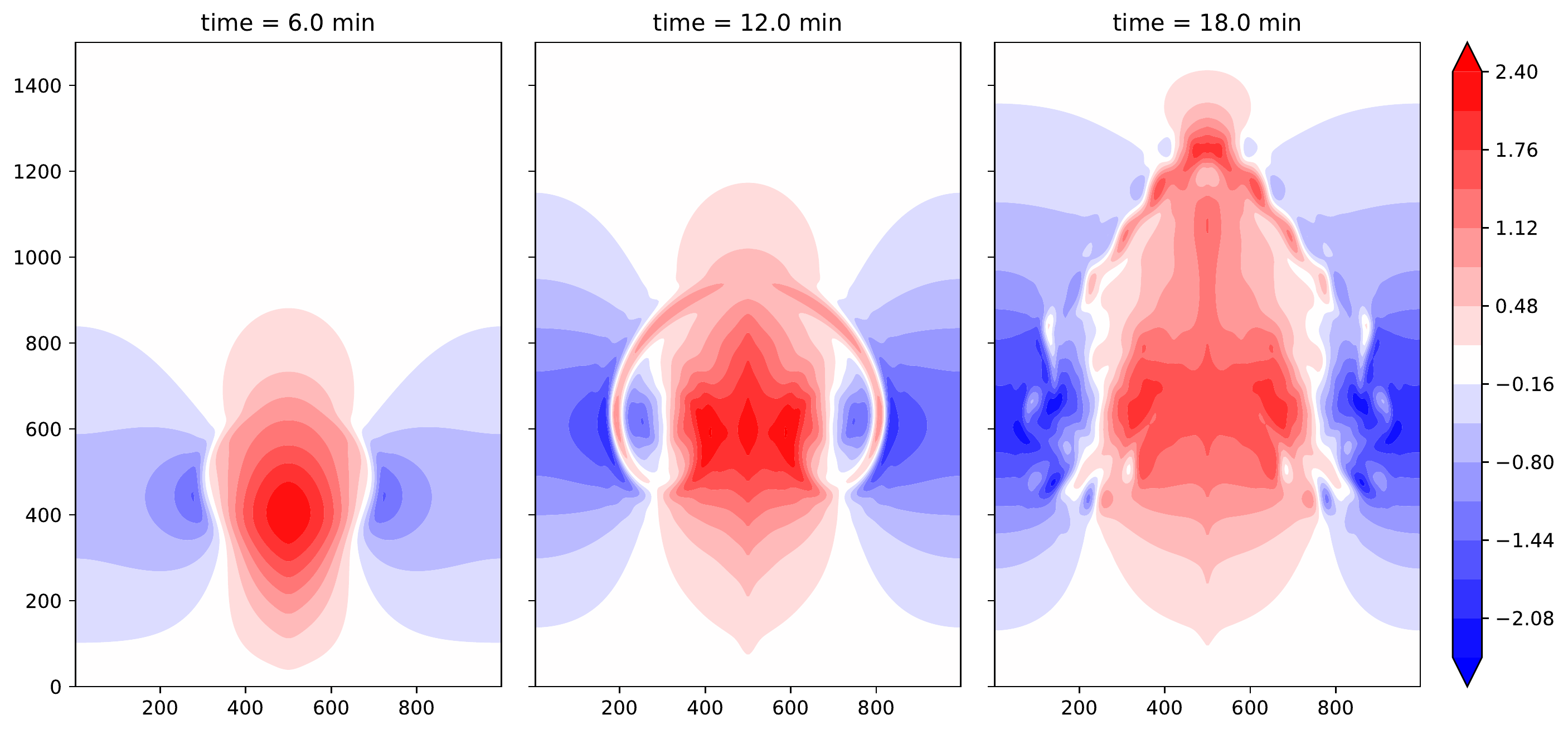}
\caption{Vertical velocities of a Gaussian-shaped rising bubble. Time is indicated in the title of each panel. Spatial resolution is 5m. Domain size is 1km by 1.5km.}
\label{fig:robert_w}
\end{figure}

The second case tests the turbulence field generated by a warm rising bubble \citep{Robert93}. The background atmospheric temperature is still an adiabat at 303.15 K and the surface pressure is 1 bar. A warm bubble, of a peak temperature excess 0.5 K, is placed near the bottom of the domain. Since the forcing is weak, the solution is more sensitive to the numerical scheme than the first case. The computational domain is 1 km wide and 1.5 km tall, with a spatial resolution of 5 m. Its evolution with time is displayed in Figure \ref{fig:robert} (potential temperature anomaly) and Figure \ref{fig:robert_w} (vertical velocity). The bubble rises and develops two rotors at both sides. Vertical velocity is positive at the arch of the rotor and negative both inside and outside. At about 18 minutes, the Kelvin-Helmholtz instability due to the vertical velocity shear destroys the smooth arch. Two rotors become turbulent and stop rising. The warmest part of the bubble concentrates at its head, protrudes out, and develops a secondary circulation. The turbulent potential temperature field at 18 minutes is identical to the reference solution calculated by \citet{Chen13} and \citet{Guerra16}. Though simple as it looks, many advanced numerical models failed to produce the correct turbulence field; some models generated more KH instabilities along the arch of the thermal bubble, and others, being too diffusive, did not produce any stability at all (e.g. \citealt{Konor14,Flyer16,Abdi17}). Passing this test indicates that our model can capture the details of the turbulence field induced by weak forcing.

\begin{figure}[ht!]
\centering
\includegraphics[scale=0.5]{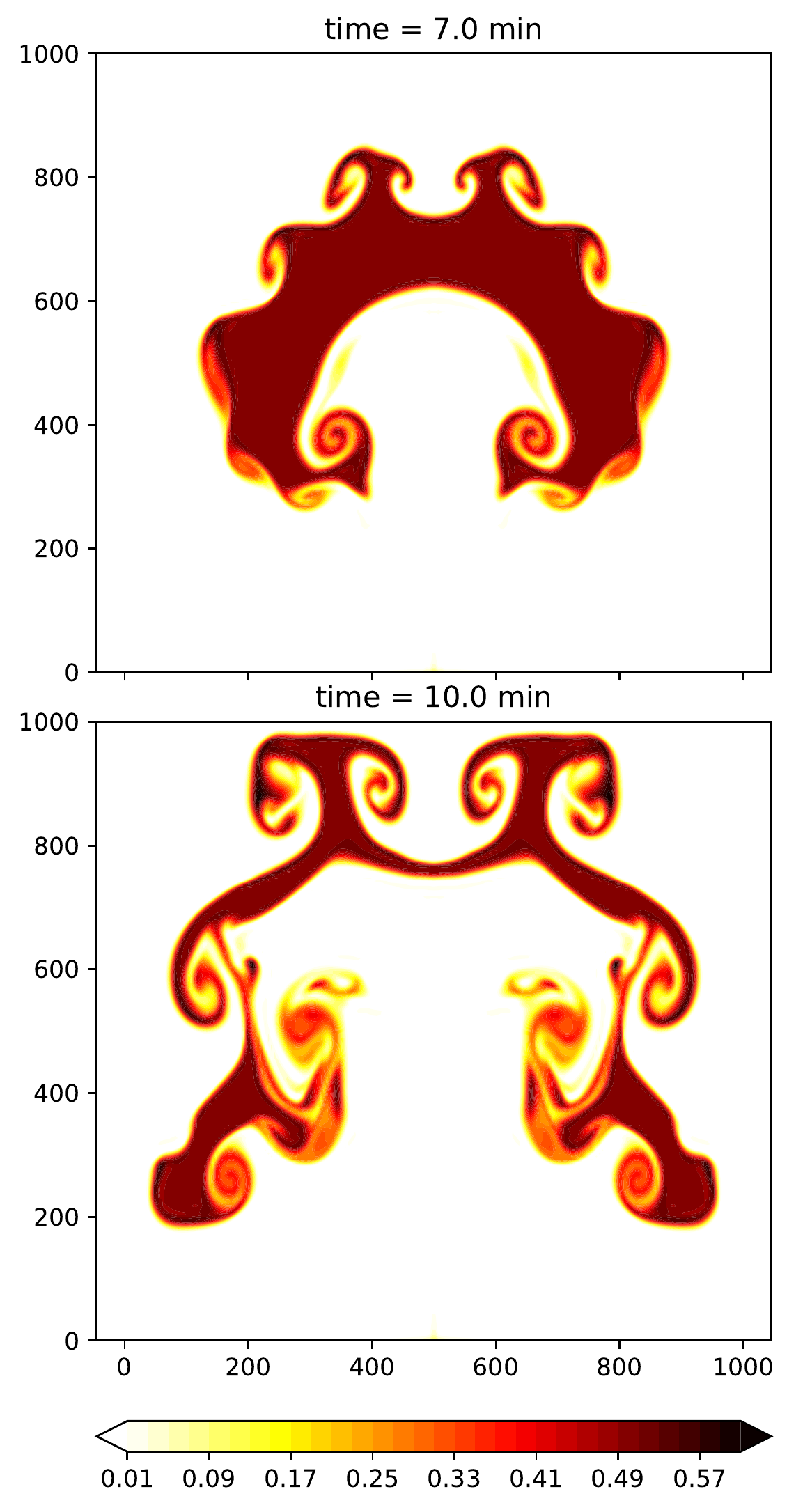}
\caption{Potential temperature anomalies of a uniform temperature rising bubble. Time is indicated in the title of each panel. Spatial resolution is 5m. Domain size is 1km by 1km.}
\label{fig:robert2}
\end{figure}

A more difficult problem arises when the rising bubble has a uniform potential temperature. Weak forcing and having a discontinuity in the temperature field make the solution very susceptible to the details of the numerical treatment. Similar to the previous one, the temperature excess is 0.5 K, and the warm bubble is evolved to 10 mins. As far as we know, there is no consensus on what the true solution is. Various authors get similar but somewhat different results. \citet{Robert93}'s solution (their Figure 2) developed one rotor at 7 mins at the shoulder of the warm bubble while we have two (Figure \ref{fig:robert2}). \citet{Chen13} did not provide a solution at 5m resolution, but the turbulence field in our result at 10 mins is very similar to theirs. 

\subsection{Bryan moist bubble}
Extending the dry simulation to a moist one with vapor and cloud, we perform the final test, which simulates a saturated bubble rising in a saturated environment. The initial temperature profile is a moist adiabat, and the total water mixing ratio, including vapor and cloud, is constant at all levels. The surface temperature and pressure are 289.85 K and 1 bar, respectively. A similar Gaussian-shaped bubble is placed near the bottom of the domain, but filled with water vapor. The bubble is warmer than the environment by less than 2 K. Thus, it contains slightly more water vapor and slightly less cloud so that it is both thermally and compositionally buoyant (background atmospheric composition is O$_2$+N$_2$). The temperature and the pressure of the bubble drop when it rises, causing water vapor to condense and latent heat released. The domain size is 10 km in the vertical and 20 km in the horizontal, with a spatial resolution of 100 m. Other details of the initial condition are described in \citet{Bryan02}. For readers who are not familiar with moist thermodynamics, we summarize the key equations and definitions in Appendix \ref{sec:thermo}.
%
%
%
%
%
%
%
%

Because we are using total energy as the thermodynamic variable, the temperature of the background atmosphere is obtained by integrating the moist adiabatic temperature gradient (see equation \ref{eqn:mgrad}). An adaptive mesh is used to assure that the background atmosphere has a uniform equivalent potential temperature profile as demanded by the test case. We tried two ways of implementing instantaneous condensation when supersaturation occurs. One is to choose a very large value of condensation rate, for example, $10^9$ s$^{-1}$. The other is to calculate the equilibrium state iteratively, a process called saturation adjustment (see Appendix \ref{sec:ssat} for the numerical scheme). We found that both ways yield the same result. 

\begin{figure}[ht!]
\centering
\includegraphics[scale=0.8]{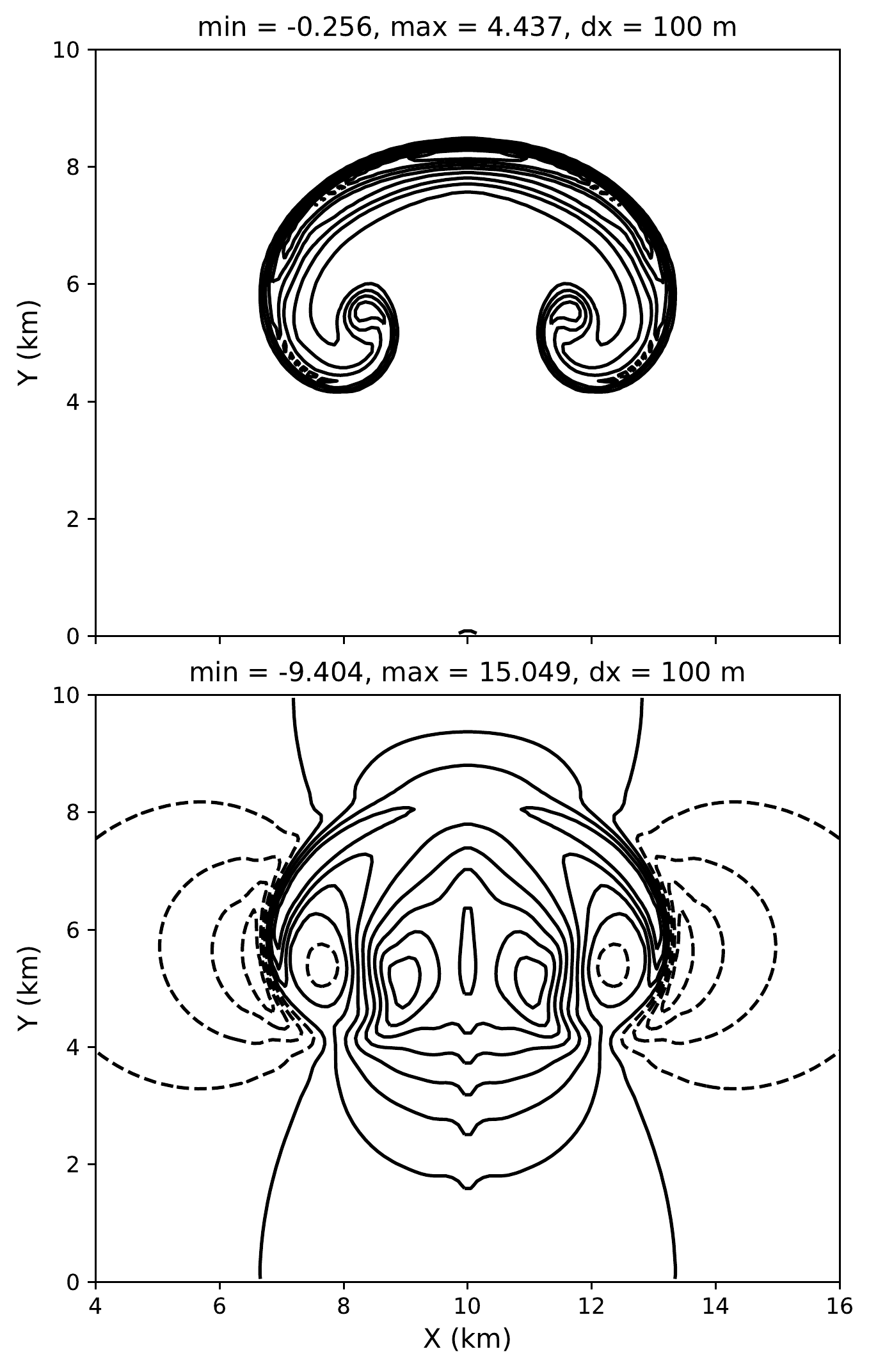}
\caption{Top panel: equivalent potential temperature anomaly, contoured every 0.5 K. Bottom panel: vertical velocity, contoured every 2 m/s. Both variables are evaluated at 1000 s. Resolution is 100 m. The minimum and maximum values are indicated in the title of each panel.}
\label{fig:bryan}
\end{figure}

Figure \ref{fig:bryan} shows the equivalent potential temperature anomaly and the vertical velocity. Our result is almost identical to the reference solution given by \citet{Bryan02} at 100m resolution, except that the maximum equivalent potential temperature anomaly (see its definition at equation \ref{eqn:theta_e}) is 4.4 K at 1000s in our simulation whereas \cite{Bryan02} got 4.09 K. We suspect that the difference is due to the difference in formulating the thermodynamic equation. Because the equivalent potential temperature is a variable representing entropy, the increase of the maximum value in an adiabatic environment may result from irreversible mixing, which cannot be captured if entropy or potential temperature is used as the prognostic variable unless all the irreversible sources of entropy production are explicitly accounted for. \cite{Pressel15} provided a list of irreversible entropy production due to mixing and precipitation process. However, fundamental difficulty arises from the estimation of the entropy generation by numerical diffusion due to the discretization of the governing equations. Although we cannot prove rigorously that the increase in the maximum equivalent potential temperature is caused by irreversible mixing, this exercise shows that using total energy as a prognostic variable obeys thermodynamic laws with less effort, and can potentially capture entropy production due to irreversible mixing, which may be an advantage over the traditional models using potential temperature as the prognostic variable. 

\section{Simulation of Jovian atmosphere in radiative-convective equilibrium } \label{sec:jupiter}
In terms of a real application to planets, we design an idealized model for the Jovian atmosphere, whose primary focus is to investigate how the internal heat is transported through the weather layer and finally radiates into the space at the optically thin level; for that reason, we are interested in the mean profiles of water, ammonia, and temperature in radiative-convective equilibrium (RCE) rather than the initial ones that trigger moist convection \citep{Stoker86,Yair95,Hueso01}. The computational domain is two-dimensional, covering 1667 km in the horizontal and 400 km in the vertical, with a spatial resolution of around 5.5 km in both directions. A doubly periodic horizontal boundary condition is assumed. The model stays one step ahead of the one-dimensional equilibrium condensation models (e.g. \citealt{Weidenschilling73}) or one-dimensional models with parameterized eddy diffusion (e.g. \citealt{Ackerman01}) by directly simulating the dynamic effects. 

We use a spatially homogeneous and temporally constant body cooling rate as a substitute for the radiative cooling process since our main focus is convection. The cooling rate is set to 1 K/day between 2 bars and 0.2 bar, which is about 100 times larger than what Jupiter emits realistically. Choosing a larger cooling rate shortens the period needed to equilibrate. Between 0.2 bar and the top boundary, there is a sponge layer absorbing kinetic energy. The total energy is not modified in the sponge layer, and therefore the damped kinetic energy turns into heat, mimicking the effect of wave breaking in the stratosphere. The sponge layer is formulated by Newtonian relaxation:

\begin{eqnarray}
\frac{\partial(\rho u)}{\partial t} & = & -\frac{\rho u}{\tau_0(p/p_0)^2} \\
\frac{\partial(\rho v)}{\partial t} & = & -\frac{\rho v}{\tau_0(p/p_0)^2},
\end{eqnarray}
where $\tau_0=10^4$ s and $p_0=0.2$ bars. The choice of the lower boundary condition matters for the dynamics since the lower boundary is the source of energy and is the place where momentum dissipates. Here, we apply a no-slip boundary condition at 200 bars to represent the damping of momentum by magnetohydrodynamic drag that acts deep in the atmosphere \citep{Grote01,Liu08a}, and a fixed temperature of 800 K to represent the internal heat source. Although the actual momentum dissipation level is much deeper than what employed here, this drag provides the principle momentum dissipation mechanism on Jupiter. A thorough discussion of the role of the lower boundary condition shall be canvassed in a dedicated study later. 

Two condensible gases, water and ammonia, as well as their clouds and precipitation are included in the model. The mass mixing ratios of water and ammonia at depth are 40 g/kg and 2.7 g/kg respectively, corresponding to a nominal water cloud base at 7 bars and ammonia cloud base at 0.7 bars. The initial temperature profile is a pseudo moist adiabat up to 0.2 bars, then isothermal above. Microphysical rate constants are largely uncertain in the Jovian atmosphere due to the unknown size distribution of clouds and precipitation. We choose nominal values of $k_1=10^9$ s$^{-1}$, $k_2=10^{-4}$ s$^{-1}$, $k_3=0$ s$^{-1}$, $k_4=10^{-2}$ s$^{-1}$, and $k_5=-20$ m/s (see Section \ref{sec:mp} for the definition of the coefficients). The extremely large value of $k_1$ manifests that condensation occurs instantaneously when vapor is saturated. The autoconversion timescale is $10^4$ s, in the middle of the range studied by \cite{Sugiyama14}. The evaporation rate and terminal velocity are roughly estimated since water and ammonia precipitation will eventually boil at 30 bars and 4 bars respectively (Figure \ref{fig:svp_tp}). 
%
%
%
%

\begin{figure}[ht!]
\centering
\includegraphics[scale=0.5]{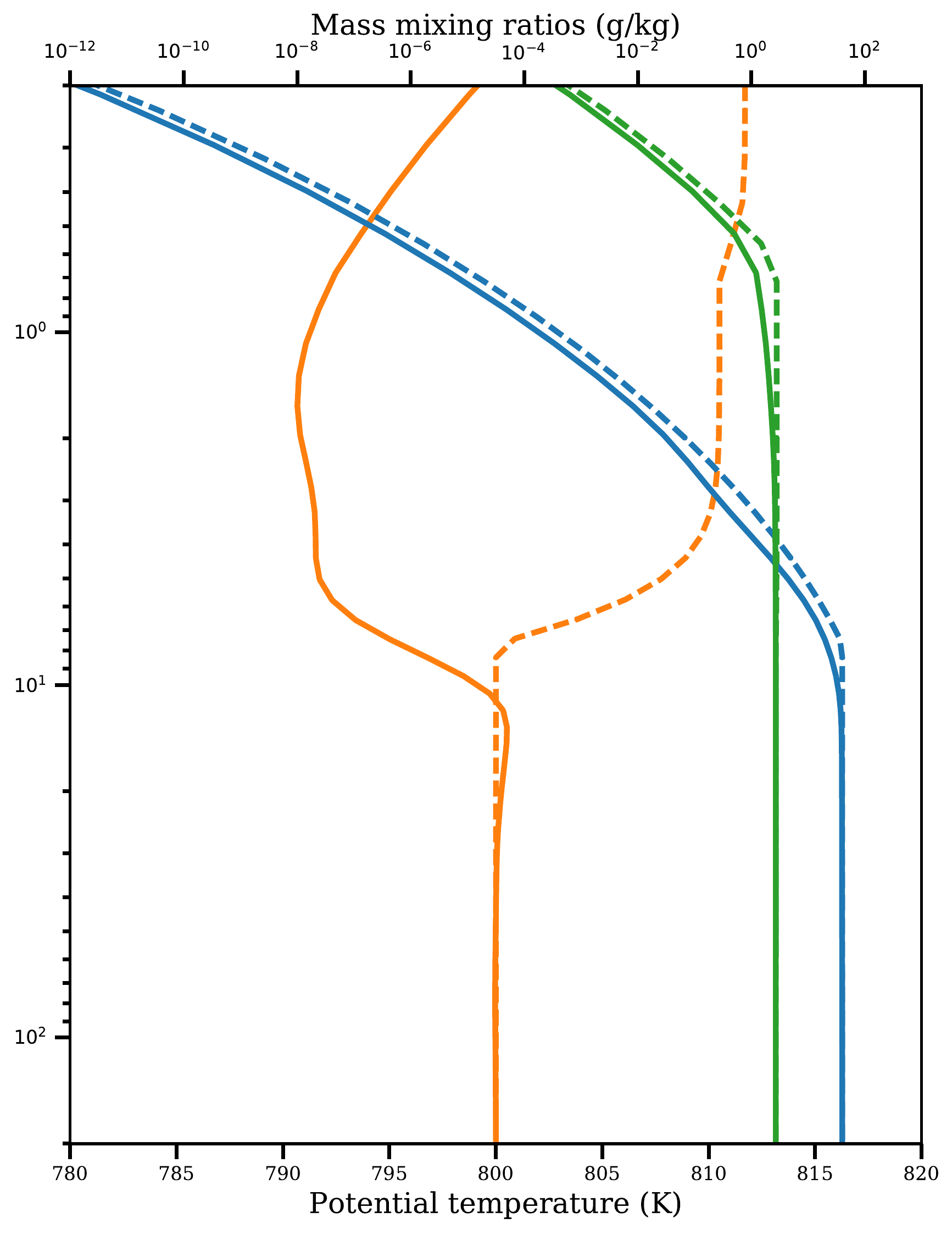}
\caption{Green lines, blue lines and orange lines show the profiles of ammonia, water and potential temperature respectively. The solid lines show the mean vertical profiles of the Jovian atmosphere in RCE. The dashed lines are the initial pseudo moist adiabatic profiles.}
\label{fig:init}
\end{figure}

\begin{figure}[ht!]
\centering
\includegraphics[scale=0.5]{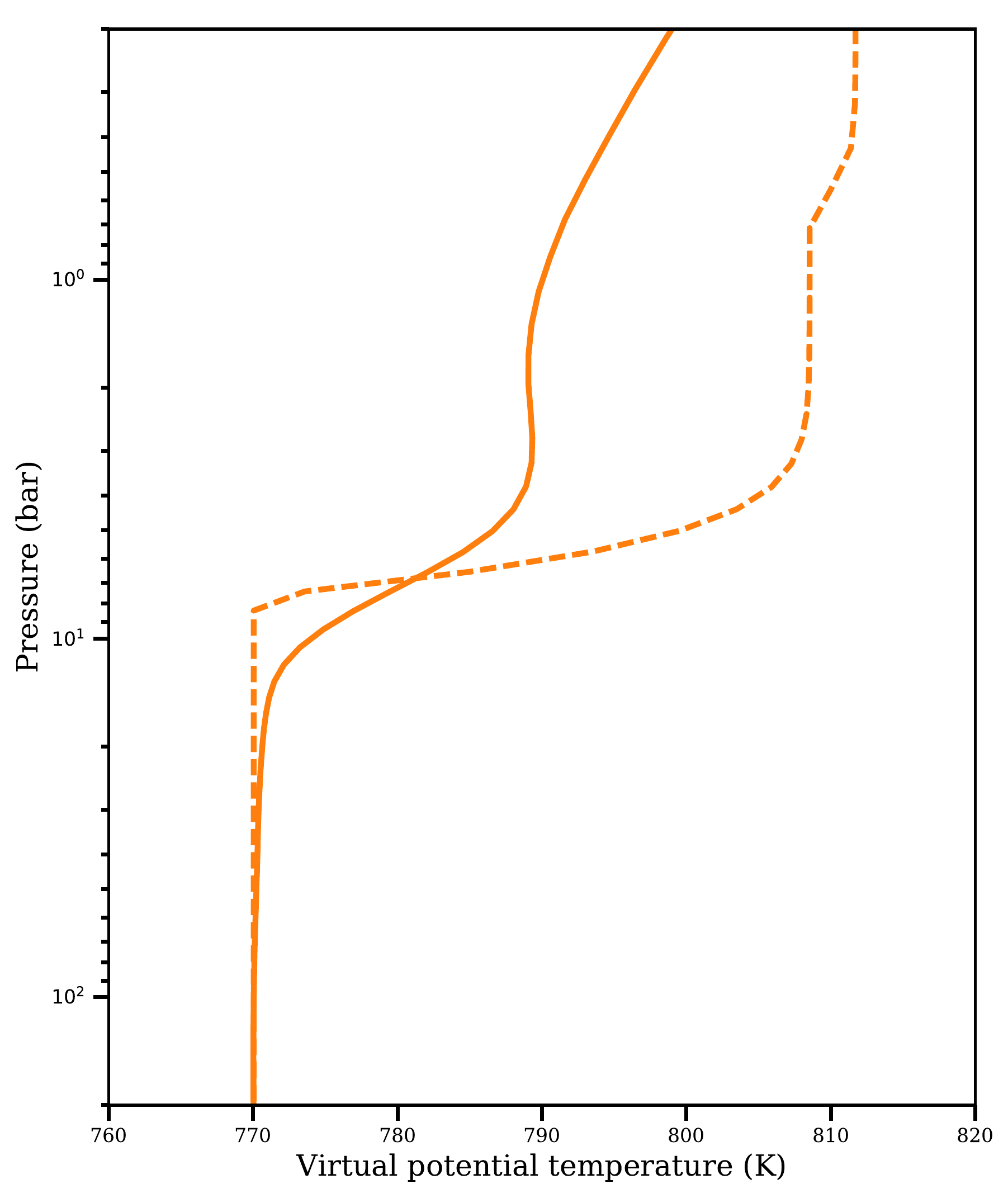}
\caption{Dashed line: virtual potential temperature profile of a pseudo moist adiabat. Solid line: mean virtual potential temperature profile of the Jovian atmosphere in RCE.}
\label{fig:thetav}
\end{figure}

The most prominent and striking feature of the Jovian atmosphere in RCE is that the mean potential temperature profile decreases with altitude between 10 bars and 5 bars (solid red line in Figure \ref{fig:init}), suggesting a superadiabatic temperature gradient. However, the seemingly unstable temperature gradient is balanced by a negative gradient of the concentration of water vapor (solid blue line in Figure \ref{fig:init}) such that the density profile is still stably stratified, as indicated by a positive gradient of the virtual potential temperature profile (Figure \ref{fig:thetav}). The amount of water and ammonia gas is also much less than their initial values, partly due to the cold atmosphere between 7 bars and 1 bar. This result contrasts the vertical profiles obtained by a one-dimensional equilibrium condensation model \citep{Weidenschilling73} that the temperature profile is a moist adiabat and that ammonia and water are well-mixed in the subcloud layer. Since the temperature at 1 bar pressure level is measured as 165 K by \cite{Lindal81}, a superadiabatic temperature gradient indicates that Jupiter's interior may be warmer than what has been assumed before based on adiabatic extrapolation. Though water vapor contributes to less than one percent of the total mass of the atmosphere in the weather layer, the condensation of water vapor significantly changes the mean molecular weight of the atmosphere, and thus the thermal stratification.

\begin{figure}[ht!]
\centering
\includegraphics[scale=0.5]{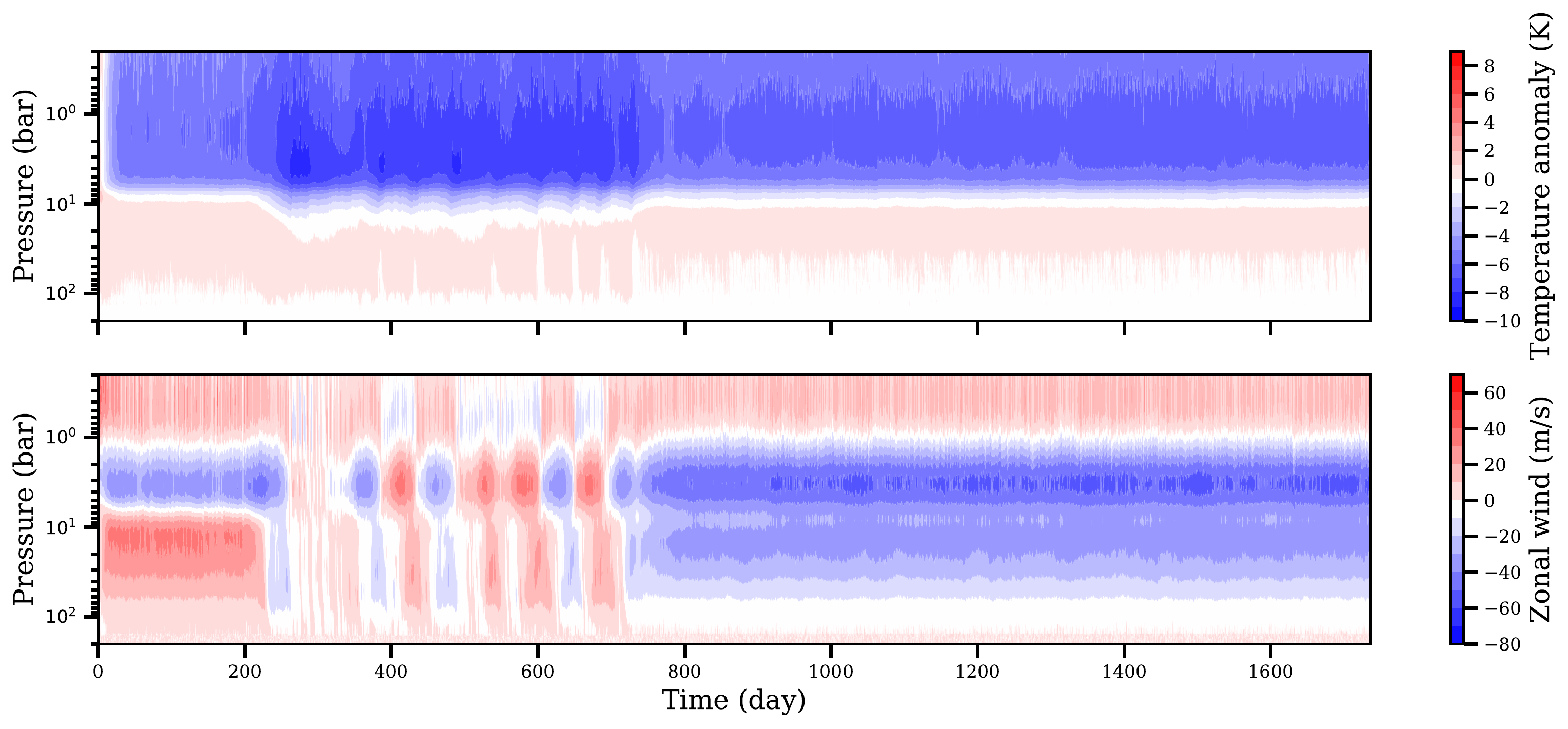}
\caption{Top panel: zonal mean temperature anomaly as a function of time with respect to the initial condition (a moist adiabat). Bottom panel: zonal mean zonal wind as a function of time.}
\label{fig:uwind}
\end{figure}

It is illuminating to see how the superadiabatic temperature gradient develops and maintains over time. We have integrated the model for over 1700 days and plotted the evolution of mean temperature anomaly (with respect to the initial temperature) and mean zonal wind in Figure \ref{fig:uwind}. The superadiabatic temperature gradient develops soon after the model started. Along with it, negative wind appears above the cloud, and positive wind appears below the cloud. The wind shear across the water cloud breaks after 200 days, likely triggered by convective instability because a large reduction in temperature above 7 bars occurs simultaneously. An unsteady state featured by alternating zonal wind lasts for about 500 days. The period of the oscillation is about 50 days, but sometimes one oscillatory phase may be skipped, for example, between day 500 and day 600. After 800 days, the atmosphere reaches a steady state. The zonal wind is strongest in the water cloud, reaching 50 m/s, and decreases with depth toward zero. The final equilibrated temperature above 7 bars is slightly warmer than what has been in the unsteady state. No matter whether the steady state is reached or not, a colder atmosphere above the water condensation level is a robust feature of the simulated Jovian atmosphere in RCE. 

\begin{figure}[ht!]
\centering
\includegraphics[scale=0.6]{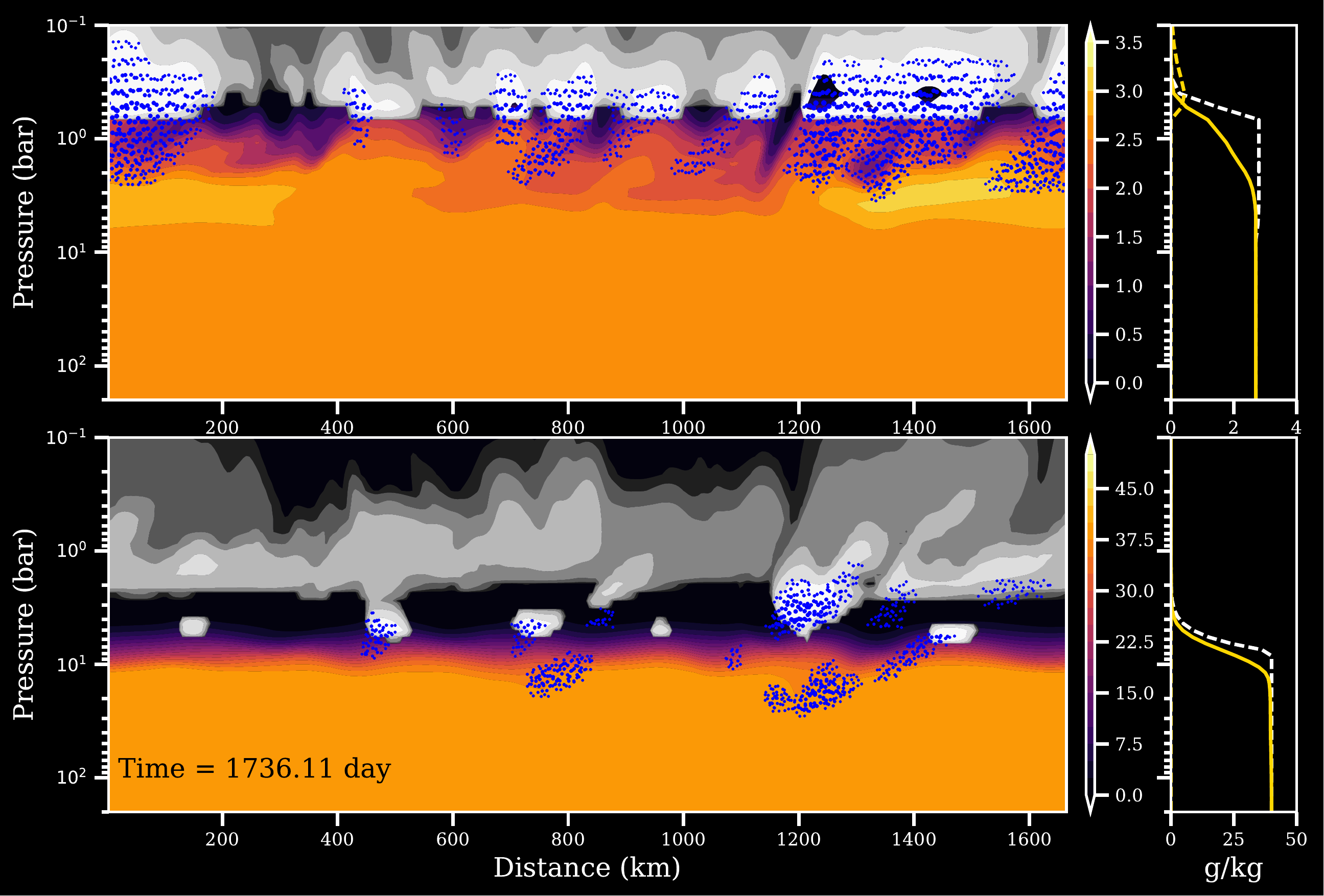}
\caption{A snapshot of the cloud structure in RCE. Top left: distribution of ammonia vapor (colored contours), cloud (white contours) and precipitation (blue dots). The contour levels of the ammonia vapor are shown in the color bar attached to the right. The contour levels of the ammonia cloud start from $10^{-3}$ g/kg and increment by a factor of $\sqrt{10}$. Top right: mean ammonia vapor profile (solid yellow), mean ammonia cloud (dashed yellow), initial ammonia profile (dashed white). Bottom: same as the top figures but for water.}
\label{fig:cloud}
\end{figure}

A snapshot of the spatial distribution of water and ammonia is shown in Figure \ref{fig:cloud}, where the vapors are plotted in colored contours, clouds in shaded contours, and precipitation in scattered dots. Dark areas in each panel indicate regions devoid of the vapor. Unlike the prediction by equilibrium condensation models, the main cloud water appears at about 2 bars, with patchy and cumulus-like cloud scattered around 5 -- 7 bars. Large cloud concentration usually coincides with clustered precipitation. Some precipitation deposits to levels deeper than 10 bars but none exists below 30 bars because of boiling. Ammonia cloud, on the other hand, forms at 0.7 bar, similar to the prediction by equilibrium condensation models, but the concentration of ammonia gas shows large variability in the subcloud layer. Some regions bear more ammonia than the deep atmosphere in mass mixing ratios, due to the re-evaporation of the ammonia precipitation, others are significantly less because compensating downdrafts carry dry air downward. On average, the concentration of the ammonia gas increases with depth and reaches its deep value at about 10 bars, where the atmosphere turns neutrally stable. Similar depletion of ammonia gas in the subcloud level was also found in the simulation result of \cite{Sugiyama14}, although the level where ammonia recovers is deeper in our simulation than in \cite{Sugiyama14}'s. 

The recent observations from the Juno spacecraft revealed the thermal emission of Jupiter's atmosphere from depth well below the water cloud \citep{Bolton17}. Yet, the interpretation of its microwave spectra is confounded by the nonuniform distribution of ammonia gas \citep{Li17,Ingersoll17}. It remains a mystery that why ammonia, an essentially chemically inert species, is depleted in the atmosphere far below its cloud condensation level. The previous study attributed the desiccation of the ammonia gas after a single convective event to the compensating downdrafts that brought dry air down from higher up \citep{Li15}. However, it is unclear how this mechanism works in a steady state in which the atmosphere is continuously heated from below and cooled from the top. This simulation demonstrates that moist convection alone can cause the ammonia gas to deplete at levels well below its nominal cloud bottom. Therefore, the effects of dynamics must be considered in modeling the vertical structure of a planet's atmosphere.

\section{Conclusion and discussion} \label{sec:conc}
In this work, we presented a new non-hydrostatic and cloud-resolving model for planets. It is developed with the goal of simulating cloud formation on both terrestrial planets and giant planets. We build this model on top of the Athena++ framework, allowing it to exploit the features provided by the Athena++ code, such as static/adaptive mesh-refine, curvilinear geometry, and dynamic task scheduling. In order to correctly and efficiently simulate atmospheric flows, we have implemented several extensions to the original Athena++ solver. They are summarized in the following.
\begin{enumerate}
\item The continuity equations are extended to vapors, clouds and precipitation.
\item Latent heat terms are added to the energy equations.
\item The equation of state is modified to represent a heterogeneous mixture of gases and clouds.
\item Microphysical processes are formulated generically such that both terrestrial and Jovian planets can use the same framework. Specifically, we discussed the formulation of condensation, precipitation and evaporation.
\item The LMARS Riemann solver is used instead of the Roe or HLLC solver for it better preserves hydrostatic balance in the vertical.
\item A fifth-order WENO reconstruction method is implemented to achieve a high order of accuracy as well as to minimize the phase error in the wave propagation.
\end{enumerate}

We pioneered the use of total energy as the prognostic variable for the thermodynamic equation while a more conventional choice is potential temperature. Potential temperature, a concept primary for interpretive purposes, becomes complicated when the heat capacity of the atmosphere varies with temperature or when multiple condensing species exist in the atmosphere as so in the Jovian atmosphere. Some authors prefer using entropy rather than potential temperature for Earth's atmosphere (e.g. \citealt{Hauf87,Pressel15}). However, fundamental difficulties arise when one wishes to account for all entropy production due to irreversible processes, including the entropy generation by numerical diffusion due to the discretization of the governing equations. Therefore, we argue that using total energy is a better choice since total energy is conserved regardless of whether the process is reversible or not. We carried out a series of numerical tests to demonstrate the veracity of the model. The simulation results are almost identical to the reference solutions in the literature.

Then, we designed an idealized simulation of the Jovian atmosphere in radiative-convective equilibrium, with water and ammonia included as condensible species. Our simulation result showed that the concentration of the ammonia gas is highly variable in the subcloud layer, with some region enriched in ammonia caused by re-evaporation and some region developed in ammonia caused by downdrafts. On average, the subcloud layer is depleted in ammonia gas and the depletion can reach 10 bars in pressure. Although, at this stage, we cannot claim that this is exclusively the cause of the observed depletion of ammonia in Jupiter's atmosphere \citep{Li17} because we have not yet considered the effect of large-scale circulation, the Coriolis force, and the three-dimensionality, we can assure that moist convection alone can cause the ammonia gas to deplete at levels well below its nominal cloud bottom. As a result, the concentration profiles resulted from the equilibrium condensation calculation should be interpreted qualitatively but not quantitatively. 

Moreover, in contrast to the assumed adiabaticity of Jupiter's atmosphere, we found that the temperature gradient is superadiabatic near the water condensation level, due to the changing of the mean molecular weight. We are certain about the mechanism but less sure on the exact magnitude because we have not yet tested the effect of the lower boundary condition and the effect of the cooling rate, which we will leave for future exploration.

We expect many future applications of the cloud-resolving model to planetary atmospheres. First, the speed of sound on Mars is about 2/3 of what it is on Earth, but the pressure fluctuation is much larger. Particularly, Mars has been shown in the MACDA dataset to have transonic jet streaks \citep{Lewis17}. Thus, our model, using the LAMRS Riemann solver, offers advantages to investigate the dynamics of the strong polar jet. Other planets with tenuous atmospheres, such as Io and Pluto, will also benefit on this point.

Second, many GCMs for giant planets have been developed with an active hydrological cycle and microphysics \citep{Palotai08,Lian10}. But a GCM cannot directly resolve moist convection due to the coarse resolution and hydrostatic assumption. Our cloud-resolving model can complement the large-scale models by devising convective parameterization schemes. Conversely, large-scale models help our simulation by providing the context and forcing that drive the cloud-resolving model.

Third, advances in the observations of exoplanets have further broadened the potential application of a cloud-resolving model. Many exoplanets' atmospheres are thought to harbor multiple layers of exotic clouds made of KCl, ZnS, MgSiO3, etc. \citep{Sing16,Morley12,Crossfield14}. Understanding their vertical structure and extend is important for interpreting the transit spectra. However, most cloud models for exoplanets' atmospheres are restricted to 1D, in which physical processes are mingled in an ad-hoc parameters, $K_{zz}$ \citep{Ackerman01,Zhang18a,Zhang18b}. There is no physical basis behind which planet should have what values of $K_{zz}$ other than it fits the observation \citep{Li14}. Able to resolve the turbulence field, a cloud-resolving model will lay a solid foundation for the physical processes governing the distribution and structure of the clouds.

\acknowledgments

We thank all the people that are working or have worked on the development of Athena++ code, and hosted it on an open source platform. It is supposed to be a software that solves astrophysical problems but we found no difficulty to extend the solver to atmospheric flows. We also thank Andrew Ingersoll, Xi Zhang, Zhaohuan Zhu, Zhihong Tan, Xianyu Tan and Kyle Pressel for stimulating discussions of the development of the model. C.L. is supported by the Juno mission.



\clearpage
\appendix

\section{List of symbols} \label{sec:symbols}
All important symbols used in the manuscript and their CGI units are collected in the following table for reference.
\startlongtable
\begin{deluxetable}{c|c|c}
\label{tab:sym}
\tablecaption{List of symbols}
\tablehead{\colhead{Notation} & \colhead{Meaning} & \colhead{Units}}
\startdata
\(\rho_d\) & density of dry air & $\mathrm{kg/m^3}$ \\
\(\rho_i\) & density of vapor $i$ & $\mathrm{kg/m^3}$ \\
\(\rho_{ij}\) & density of the condensate $j$ associated with the vapor $i$ & $\mathrm{kg/m^3}$ \\
\(\rho\) & total density, \(\rho=\rho_d+\sum_i\rho_i+\sum_{i,j}\rho_{ij}\) & $\mathrm{kg/m^3}$ \\
\(q_d\) & mass mixing ratio of dry air, \(q_d=\frac{\rho_d}{\rho}\) & 1 \\
\(q_i\) & mass mixing ratio (specific humidity) of vapor $i$, \(q_i=\frac{\rho_i}{\rho}\) & 1 \\
\(q_{ij}\) & mass mixing ratio of condensate $(i,j)$, \(q_{ij}=\frac{\rho_{ij}}{\rho}\) & 1 \\
\(q^*_{ij}\) & saturation mass mixing ratio of vapor $i$ over condensate $(i,j)$ & 1 \\
\(g\) & gravitational acceleration & $\mathrm{m^2/s}$ \\
\(x, y\) & horizontal coordinates & m \\
\(z\) & vertical coordinate & m \\
\(u,v\) & horizontal velocities & $\mathrm{m/s}$ \\
\(w\) & vertical velocity & $\mathrm{m/s}$ \\
\(w^t_{ij}\) & sedimentation velocity of condensate $(i,j)$ & $\mathrm{m/s}$ \\
\(p_d\) & partial pressure of dry air & $\mathrm{pa}$ \\
\(p_i\) & partial pressure of vapor $i$ & $\mathrm{pa}$ \\
\(p_{ij}^*\) & saturation vapor pressure of vapor $i$ over condensate $(i,j)$ & $\mathrm{pa}$ \\
\(p\) & total pressure, \(p=p_d+\sum_i p_i\) & $\mathrm{pa}$ \\
\(e\) & total specific energy (not including gravitational potential) & $\mathrm{J/kg}$ \\
\(\mathbf{X}\) & primitive variables $\mathbf{X}=(\rho,q_i,q_{ij},u,v,w,p)^T$ &  \\
\(\mathbf{Y}\) & conserved variables $\mathbf{Y}=(\rho_d,\rho_i,\rho_{ij},\rho u,\rho v,\rho w,\rho e)^T$ &  \\
\(c_{v,d}\) & specific heat capacity of dry air at constant volume  & $\mathrm{J/(kg\,K)}$ \\
\(c_{p,d}\) & specific heat capacity of dry air at constant pressure & $\mathrm{J/(kg\,K)}$ \\
\(c_{v,i}\) & specific heat capacity of vapor $i$ at constant volume & $\mathrm{J/(kg\,K)}$ \\
\(c_{p,i}\) & specific heat capacity of vapor $i$ at constant pressure & $\mathrm{J/(kg\,K)}$ \\
\(c_{ij}\) & specific heat capacity of condensate $(i,j)$ & $\mathrm{J/(kg\,K)}$ \\
\(\Delta c_{ij}\) & difference of specific heat capacity between condensate and vapor, $\Delta c_{ij} = c_{ij}-c_{p,i}$ & $\mathrm{J/(kg\,K)}$ \\
\(L_{ij}\) & latent heat of forming condensate $(i,j)$ & $\mathrm{J/kg}$ \\
\(L^r_{ij}\) & latent heat of forming condensate $(i,j)$ at reference temperature $T^r$ & $\mathrm{J/kg}$ \\
\(T^r\) & reference temperature & $\mathrm{K}$ \\
\(p^r_{ij}\) & saturation vapor pressure of vapor $i$ over condensate $(i,j)$ at reference temperature $T^r$ & $\mathrm{pa}$ \\
\(\mu_i\) & chemical potential of vapor $i$ & $\mathrm{J/kg}$ \\
\(\mu_{ij}\) & chemical potential of condensate $(i,j)$ & $\mathrm{J/kg}$ \\
\(m_d\) & molecular weight of dry air & $\mathrm{kg/mol}$ \\
\(m_i\) & molecular weight of vapor $i$ & $\mathrm{kg/mol}$ \\
\(\hat{R}\) & ideal gas constant, \(\hat{R}=8.3144598\) J/(mol K) & $\mathrm{J/(mol\,K)}$ \\
\(R_d\) & ideal gas constant of dry air, \(R_d=\hat{R}/m_d\) & $\mathrm{J/(kg\,K)}$ \\
\(R_i\) & ideal gas constant of vapor $i$, \(R_i=\hat{R}/m_i\) & $\mathrm{J/(kg\,K)}$ \\
\(\epsilon_i\) & ratio of molecular weights between vapor and dry air, \(\epsilon_i=m_i/m_d\) & 1 \\
\(\sigma_{v,i}\) & ratio of specific heats between vapor and dry air, $\sigma_{v,i}=c_{v,i}/c_{v,d}$ & 1 \\
\(\sigma_{p,i}\) & ratio of specific heats between vapor and dry air,
$\sigma_{p,i}=c_{p,i}/c_{p,d}$ & 1 \\
\(\sigma_{v,ij}\) & ratio of specific heats between condensate and dry air, \(\sigma_{v,ij}=c_{ij}/c_{v,d}\) & 1 \\
\(\sigma_{p,ij}\) & ratio of specific heats between condensate and dry air, \(\sigma_{p,ij}=c_{ij}/c_{p,d}\) & 1 \\
\(\gamma\) & adiabatic index & 1 \\
\(\gamma_d\) & adiabatic index of dry air, \(\gamma_d=c_{p,d}/c_{v,d}\) & 1 \\
\(T\) & temperature & $\mathrm{K}$ \\
\(T_v\) & virtual temperature, $T_v = T(1+\sum_i q_i(1/\epsilon_i-1)-\sum_{i,j} q_{ij})$ & $\mathrm{K}$ \\
\(\theta\) & potential temperature & $\mathrm{K}$ \\
\(\theta_v\) & virtual potential temperature, $\theta_v=\theta(1+\sum_i q_i(1/\epsilon_i-1)-\sum_{i,j} q_{ij})$ & $\mathrm{K}$ \\
\enddata
\end{deluxetable}

\section{Conversion between primitive variables and conserved variables} \label{sec:prim}
The primitive variables are :

\begin{equation}
\mathbf{X}=(\rho,q_i,q_{ij},u,v,w,p)^T
\end{equation}
The conserved variables are:

\begin{equation}
\mathbf{Y}=(\rho_d,\rho_i,\rho_{ij},\rho u,\rho v,\rho w,\rho e)^T
\end{equation}
Converting the primitive variables to the conserved variables takes two steps. The first step converts specific humidities to densities and velocities to momentums.

\begin{eqnarray}
\rho_i & = & \rho\cdot q_i \\
\rho_{ij} & = & \rho\cdot q_{ij} \\
\rho_d & = & \rho - \sum_i \rho_i - \sum_{i,j} \rho_{ij} \\
\rho u & = & \rho\cdot u \\
\rho v & = & \rho\cdot v \\
\rho w & = & \rho\cdot w
\end{eqnarray}
The second step converts total pressure to total energy. Total pressure is the sum of the partial pressure of each gaseous component:

\begin{equation}
\begin{aligned}
p & = p_d + \sum_i p_i \\
  & = \rho_d R_d T + \sum_i \rho_i R_i T \\
  & = (\rho_d R_d + \sum_i \rho_i R_i)T \label{eqn:p} \\
\end{aligned}
\end{equation}
Total energy includes internal energy, kinetic energy, and chemical potentials:

\begin{equation}
\begin{aligned}
\rho e & =  \rho_d c_{v,d}T + \sum_i\rho_i c_{v,i}T + \sum_{i,j}\rho_{ij} c_{ij}T
	   + \frac{1}{2}\rho(u^2+v^2+w^2)+\sum_i\mu_i\rho_i+\sum_{i,j}\mu_{ij}\rho_{ij} \\
       & = (\rho_d c_{v,d} + \sum_i\rho_i c_{v,i} + \sum_{i,j}\rho_{ij} c_{ij})T
       + K + A, \label{eqn:te}\\
\end{aligned}
\end{equation}
where $K=\frac{1}{2}\rho(u^2+v^2+w^2)$ is the kinetic energy and $A=\sum_i\mu_i\rho_i+\sum_{i,j}\mu_{ij}\rho_{ij}$ is the chemical potential. Substituting equation (\ref{eqn:p}) into equation (\ref{eqn:te}) to replace the temperature gives:

\begin{equation}
\begin{aligned}
\rho e & = \frac{\rho_d c_{v,d} + \ \rho_i c_{v,i} + \sum_{i,j}\rho_{ij} c_{ij}}{\rho_d R_d + \sum_i \rho_i R_i}p + K + A \\
  & = \frac{c_{v,d}}{R_d}\frac{\rho_d + \sum_i\rho_i \sigma_{v,i} + \sum_{i,j}\rho_{ij} c_{ij}}{\rho_d + \sum_i \rho_i/\epsilon_i}p + K + A \\
  & = \frac{c_{v,d}}{R_d}\frac{1 + \sum_i q_i (\sigma_{v,i}-1) + \sum_{i,j}q_{ij} (\sigma_{v,ij}-1)}{1 + \sum_i q_i(1/\epsilon_i-1)-\sum_{i,j} q_{ij}}p + K + A \\
  & = \frac{1}{\gamma_d-1}\frac{1 + \sum_i q_i (\sigma_{v,i}-1) + \sum_{i,j}q_{ij} (\sigma_{v,ij}-1)}{1 + \sum_i q_i(1/\epsilon_i-1)-\sum_{i,j} q_{ij}}p + K + A
\end{aligned}
\end{equation}

Converting the conserved variables to the primitive variables is the reverse of the previous steps. First, the specific humidities are obtained by letting:
\begin{eqnarray}
\rho & = & \rho_d+\sum_i\rho_i+\sum_{i,j}\rho_{ij} \\
q_i & = & \rho_i/\rho \\
q_{ij} & = & \rho_{ij}/\rho \\
u & = & \rho u/\rho \\
v & = & \rho v/\rho \\
w & = & \rho w/\rho
\end{eqnarray}
Define internal energy as:

\begin{equation}
\begin{aligned}
U & = \frac{1}{\gamma_d-1}\frac{1 + \sum_i q_i (\sigma_{v,i}-1) + \sum_{i,j} q_{ij} (\sigma_{v,ij}-1)}{1 + \sum_i q_i(1/\epsilon_i-1)-\sum_{i,j} q_{ij}}p \\
   & = \rho e - K - A
\end{aligned}
\end{equation}
Then the formula for pressure is:

\begin{equation}
p = (\gamma_d-1)\frac{1 + \sum_i q_i(1/\epsilon_i-1)-\sum_{i,j} q_{ij}}{1 + \sum_i q_i (\sigma_{v,i}-1) + \sum_{i,j} q_{ij} (\sigma_{v,ij}-1)}\cdot U \label{eqn:p2}
\end{equation}
Now we can define an effective adiabatic index $\gamma$ for an heterogeneous air parcel:

\begin{equation}
\gamma = 1+(\gamma_d-1)\frac{1 + \sum_i q_i(1/\epsilon_i-1)-\sum_{i,j}q_{ij}}{1 + \sum_i q_i (\sigma_{v,i}-1) + \sum_{i,j} q_{ij} (\sigma_{v,ij}-1)}
\end{equation}
It can be shown that this definition is consistent with the definition of potential temperature in equation (\ref{eqn:theta}). In the special case in which there are no clouds ($q_{c_i}=0$) and the adiabatic indices of all vapors equal that of dry air:

\begin{equation}
\sigma_{v,i} = \frac{1}{\epsilon_i}
\end{equation}
The adiabatic index of the air mixture equals that of the dry air: $\gamma = \gamma_d$, and equation (\ref{eqn:p2}) reduces to the equation of state of a single-component ideal gas regardless of the differences in molecular weight:

\begin{equation}
p = (\gamma-1)\cdot U
\end{equation}

\section{Thermodynamic formulas} \label{sec:thermo}
In this section, we derive the expression of several important thermodynamic variables. Some of the expressions are available in a standard textbook (e.g. \citealt{Emanuel94}). But often, these expressions only consider one condensing species, water vapor. We extend them to multiple condensing species and emphasize on the quantities that will be used to diagnose or initialize the model.
\subsection{Moist static energy}
A conserved quantity during adiabatic displacement in which the pressure change is hydrostatic is moist static energy $h_m$. The generalized expression is:

\begin{equation}
\begin{aligned}
h_m & = \big[c_{p,d}q_d + \sum_i c_{p,i}(q_i+\sum_j q_{ij})\big]T-\sum_{i,j}L_{ij}q_{ij}+gz \\
  & = c_{p,d}T\big[1+\sum_i(q_i+\sum_j q_{ij})(\sigma_{p,i}-1)\big]-\sum_{i,j}L_{ij}q_{ij}+gz \label{eqn:mse}
\end{aligned}
\end{equation}
If the atmosphere is at rest, using equations (\ref{eqn:energy}), (\ref{eqn:latent}) and (\ref{eqn:mse}), it is easy to verify that:

\begin{equation}
h_m = e + R_d T_v + gz
\end{equation}

\subsection{Moist adiabatic temperature gradient}
The moist adiabatic temperature gradient in pressure coordinate for an atmosphere with multiple condensing species is derived in \citet{Li18}, which reads:

\begin{equation}
\Big(\frac{d\ln T}{d\ln p}\Big)_{moist}=\frac{(1+\sum_ir_i/\epsilon_i)(1+\sum_ir_i/\epsilon_i\frac{L_{ij}}{R_i T})}{\frac{c_{p,d}}{R_d}(1+\sum_ir_i\sigma_{p,i}+\sum_{i,j}r_{ij}\sigma_{p,ij})+\sum_ir_i/\epsilon_i(\frac{L_{ij}}{R_i T})^2+(\sum_ir_i/\epsilon_i\frac{L_{ij}}{R_i T})^2}, \label{eqn:mgrad}
\end{equation}
where $r_i=q_i/q_d$ and $r_{ij}=q_{ij}/q_d$ are the mass mixing ratios of vapor $i$ and cloud $(i,j)$ with respect to the dry air respectively. Using the hydrostatic balance and the equation of state (equation \ref{eqn:eos}), the temperature gradient in height coordinate is:

\begin{equation}
\frac{dT}{dz}=-\frac{d\ln T}{d\ln p}\frac{g}{R_d}\frac{T}{T_v} \label{eqn:dtdz}
\end{equation}
Either equation (\ref{eqn:mse}) or equation (\ref{eqn:dtdz}) can be used to initialize a background atmosphere that is neutral to moist convection. In a model that uses height coordinate, directly integrating equation (\ref{eqn:dtdz}) is often easier than an iterative approach to find a solution that satisfies the conservation of moist static energy. Refinement in the grid size is sometimes necessary in order to achieve high accuracy, especially in the grid where condensation occurs. The pros and cons of different approaches are discussed in \cite{Li18}.

\subsection{Potential temperature}
The potential temperature is the temperature that an air parcel would have when it is adiabatically displaced to the reference pressure without having phase change. Let the latent heat terms ($L_{ij}$) to be zero, equation (\ref{eqn:mgrad}) reduces to the dry adiabatic temperature gradient:

\begin{equation}
\Big(\frac{d \ln T}{d \ln p}\Big)_{dry}=\frac{R_d}{c_{p,d}}\cdot
\frac{1+\sum_i q_i(1/\epsilon_i-1)-\sum_{i,j} q_{ij}}{1+\sum_i q_i(\sigma_{p,i}-1)+\sum_{i,j} q_{ij}(\sigma_{p,ij}-1)} \label{eqn:dgrad}
\end{equation}
Therefore, the potential temperature is obtained by integrating equation (\ref{eqn:dgrad}):

\begin{equation}
\begin{aligned}
\theta & = T(\frac{p_0}{p})^{\chi} \\
\chi & = \frac{R_d}{c_{p,d}}\cdot
\frac{1+\sum_i q_i(1/\epsilon_i-1)-\sum_{i,j} q_{ij}}{1+\sum_i q_i(\sigma_{p,i}-1)+\sum_{i,j} q_{ij}(\sigma_{p,ij}-1)} \label{eqn:theta}
\end{aligned}
\end{equation}
Similar to the definition of virtual temperature, the virtual potential temperature is defined as:

\begin{equation}
\theta_v = \theta\big[1+\sum_i q_i(1/\epsilon_i-1)-\sum_{i,j} q_{ij}\big]
\end{equation}
The gradient of the virtual potential temperature is approximately related to the gravity wave frequency, the Brunt-V\"ais\"al\"a frequency:

\begin{equation}
N^2\approx\frac{g}{\theta_v}\frac{\partial \theta_v}{\partial z} \label{eqn:thetav}
\end{equation}

\subsection{Equivalent potential temperature}
The equivalent potential temperature is a measurement of entropy taken the latent heat into account. It is conserved during adiabatic and reversible phase change. In a saturated atmosphere with a single condensible species, the formula of equivalent potential temperature is given by equation 4.5.11 in \citet{Emanuel94}:

\begin{equation}
\label{eqn:theta_e}
\theta_e = T\Big(\frac{p_0}{p_d}\Big)^{R_d q_d/(c_{p,d}q_d+c_{ij} q_t)}\exp\Big(\frac{L_{ij} q_i}{(c_{p,d}q_d+c_{ij}q_t)T}\Big),
\end{equation}
where $q_t=q_i+q_{ij}$ is the total mass mixing ratio of the vapor and the cloud. Note that this expression is only valid for a single condensible species and a single condensate. So, the expression is rarely used in our model but is described for comparison purpose because one of the test case, the Bryan moist bubble, uses equivalent potential temperature as a diagnostic variable.

\section{Saturation adjustment scheme} \label{sec:ssat}
It is easier to work with molar mixing ratios rather than mass mixing ratios when dealing with thermodynamics. Therefore, in this section, we will work with molar quantities. A variable with a hat symbol ($\hat{}$) on its top represents the molar equivalence of the usual meaning defined in Table \ref{tab:sym}. Thus, the thermodynamic state variables are:

\begin{equation}
\mathbf{\Theta}=(T,\hat{q}_i,\hat{q}_{ij},u,v,w,p)^T
\end{equation}
The following two equations relate the mass and molar representations:

\begin{eqnarray}
q_i & = & \frac{\epsilon_i \hat{q}_i}{1+\sum_i(\hat{q}_i+\sum_j\hat{q}_{ij})(\epsilon_i-1)} \\
\hat{q}_i & = & \frac{q_i/\epsilon_i}{1+\sum_i(q_i+\sum_jq_{ij})(1/\epsilon_i-1)}
\end{eqnarray}

Vapor condenses to cloud when its partial pressure exceeds the saturation vapor pressure. In a numerical model that uses control volume, condensation process happens under the condition of constant volume and constant total energy instead of the usual assumption of isobaric process. The saturation adjustment scheme efficiently calculates the transfer of mass from vapor to cloud or vice versa that would maintain thermodynamic equilibrium between vapor and cloud. There are a number of ways to calculate the saturation vapor pressure including using a look-up table or using an empirical formula such as the Antoine equation. We directly integrate the Clausius-Clapeyron relation because we will use the gradient of the saturation curve, which is exactly the Clausius-Clapeyron relation, to derive the adjustment scheme. The Clausius-Clapeyron relation is:

\begin{equation}
\frac{d \ln p_{ij}^*(T)}{dT}=\frac{L_{ij}(T)}{R_i T^2}=\frac{L^r_{ij}-\Delta c_{ij}(T-T^r)}{R_i T^2}
\end{equation}
We have used equation (\ref{eqn:latent}) to correct the temperature dependence of the latent heat. Integrating the above Clausius-Clapeyron relation gives the saturation vapor pressure:

\begin{equation}
p_{ij}^*=p^r_{ij}\exp[\beta_{ij}(1-1/t)-\delta_{ij} \ln t],
\end{equation}
where $p^r_{ij}$ is the saturation vapor pressure at reference temperature $T^r$, and $t,\beta_{ij},\delta_{ij}$ are non-dimensional parameters:
\begin{eqnarray}
t & = & T/T^r \\
\beta_{ij} & = & \frac{L^r_{ij}+\Delta c_{ij} T^r}{R_i T^r} = \frac{\mu_i-\mu_{ij}}{R_i T_{3i}} \\
\delta_{ij} & = & \frac{\Delta c_{ij}}{R_i}
\end{eqnarray}

Assuming vapor $i$ and its cloud $(i,j)$ undergo a phase change. The transfer of substance in molar amount is $\Delta \hat{q}$. The partial pressure of the vapor before is:

\begin{equation}
p_i=\frac{\hat{q}_i}{\hat{q}_d+\hat{q}_i+\sum_{j\neq i}\hat{q}_j}p=\frac{\hat{q}_i}{\hat{q}_i+g}p, 
\end{equation}
where $g=\hat{q}_d+\sum_{j\neq i}\hat{q}_j$ is the sum of the molar mixing ratios of other gases. Ignoring the change of pressure for now, after condensation, the temperature changes to $T+\Delta T$ and the partial pressure becomes the saturation vapor pressure:

\begin{equation}
p_{ij}^*(T+\Delta T)=\frac{\hat{q}_i+\Delta \hat{q}}{\hat{q}_i+\Delta \hat{q} +g}p
\end{equation}
Let $s(T)=p_{ij}^*(T)/p$, then:

\begin{equation}
\Delta \hat{q}=\hat{q}_i-\frac{s(T+\Delta T)}{1-s(T+\Delta T)}g
\end{equation}
Expand the above equation near $T$ and keep the first order term:

\begin{equation}
\label{eqn:dx}
\Delta \hat{q}=\hat{q}_i-\big[\frac{s(T)}{1-s(T)}+\frac{s'(T)}{[1-s(T)]^2}\Delta T\big]g
\end{equation}
Using the Clausius-Clapeyron relation:

\begin{equation}
\begin{aligned}
\label{eqn:sp}
\frac{d\ln s(T)}{dT} & = \frac{d\ln p_{ij}^*(T)}{dT} = \frac{L_{ij}(T)}{R_i T^2} = \frac{\beta_{ij}/t-\delta_{ij}}{T} \\
s'(T) & = s(T)\frac{\beta_{ij}/t-\delta_{ij}}{T}
\end{aligned}
\end{equation}
Let $\alpha=\Delta T/(T\Delta \hat{q})$ during the phase change. Substituting $s'(T)$ in equation (\ref{eqn:dx}) with equation (\ref{eqn:sp}) and $\alpha$ gives:

\begin{equation}
\label{eqn:dxf}
\Delta \hat{q}=\frac{\hat{q}_i-\frac{s(T)}{1-s(T)}g}{1+\frac{s(T)}{[1-s(T)]^2}(\beta_{ij}/t - \delta_{ij})\alpha g}
\end{equation}
Under isobaric process, enthalpy $h$ is conserved:

\begin{equation}
\label{eqn:h}
h =\big[\hat{q}_d \hat{c}_{p,d}+\sum_i(\hat{q}_i+\sum_k\hat{q}_{ik})\hat{c}_{p,i})\big]T-\hat{L}_{ij} \hat{q}_{ij}=constant
\end{equation}
Therefore,

\begin{equation}
\begin{aligned}
\alpha & \approx \frac{\hat{L}_{ij}}{\hat{c}_{p,d}\big[1+\sum_i (\hat{q}_i+\sum_k\hat{q}_{ik})(\hat{\sigma}_{p,i}-1)\big]} \\
	   & = \frac{\gamma_d-1}{\gamma_d}
       		\frac{\beta_{ij}/t - \delta_{ij}}{1+\sum_i(\hat{q}_i+\sum_k\hat{q}_{ik})(\hat{\sigma}_{p,i}-1)}
\end{aligned}
\end{equation}
Under isochoric process, energy is conserved:

\begin{equation}
\begin{aligned}
\label{eqn:u}
u & = h -pv \\
  & = \big[\hat{q}_d \hat{c}_{vd}+\sum_i (\hat{q}_i+\sum_k\hat{q}_{ik})\hat{c}_{v,i}\big]T-(\hat{L}_{ij}-\hat{R}T) \hat{q}_{ij}=constant
\end{aligned}
\end{equation}
Therefore,

\begin{equation}
\begin{aligned}
\alpha & \approx \frac{\hat{L}_i-\hat{R}T}{\hat{c}_{v,d}\big[1+\sum_i (\hat{q}_i+\sum_k\hat{q}_{ik})(\hat{\sigma}_{v,i}-1)\big]} \\
	   & = (\gamma_d-1)\frac{\beta_{ij}/t-\delta_{ij}-1}{1+\sum_i (\hat{q}_i+\sum_k\hat{q}_{ik})(\hat{\sigma}_{v,i}-1)}
\end{aligned}
\end{equation}

Iteration is needed in the presence of multiple condensing species. Starting from each iteration, equation (\ref{eqn:dxf}) is used to calculate the projected mass transfer. The value of $\Delta \hat{q}$ must be limited to ensure the positiveness of all substances. Then either equation (\ref{eqn:h}) or equation (\ref{eqn:u}) is used to update temperature depending on whether it is an isobaric or isochoric process, i.e. whether the vertical coordinate is pressure or height. If the phase change happens at constant volume, pressure needs to be updated accordingly:

\begin{equation}
\begin{aligned}
p & = \rho_d R_d T + \sum_i \rho_i R_i T \\
  & = \rho R_d T \frac{\hat{q}_d + \sum_i \hat{q}_i}{1 + \sum_i(\hat{q}_i+\sum_j\hat{q}_{ij})(\epsilon_i - 1)} \\
  & = \rho R_d T \frac{1-\sum_{i,j}\hat{q}_{ij}}{1 + \sum_i(\hat{q}_i+\sum_j\hat{q}_{ij})(\epsilon_i - 1)} \\
  & = \rho R_d T_v, 
\end{aligned}
\end{equation}
where

\begin{equation}
T_v = T \frac{1-\sum_{i,j}\hat{q}_{ij}}{1 + \sum_i(\hat{q}_i+\sum_j\hat{q}_{ij})(\epsilon_i - 1)}, 
\end{equation}
is another expression of virtual temperature using molar mixing ratios (c.f. equation \ref{eqn:tv}). Next iteration begins with the updated temperature, pressure and abundances until the solution converges. Because the first derivative of the saturation vapor pressure is taken care of, a quadratic convergence is expected. In real applications, we find that the iteration converges to machine precision in less than 5 iterations.

\clearpage




\bibliography{b.General}





\end{document}